\documentclass{aa}
\usepackage[usenames,dvipsnames]{xcolor}
\usepackage[T1]{fontenc}
\usepackage{natbib}
\usepackage{csquotes}
\usepackage{xspace}
\usepackage{txfonts}
\usepackage{amsmath}
\usepackage{threeparttable}
\usepackage{graphicx}
\usepackage{hyperref}
\usepackage[separate-uncertainty=true]{siunitx}
\usepackage{tikz}
\usepackage{subcaption}
\usepackage{placeins}

\def\max{\mathrm{max}}

\def\Hz{\hbox{Hz}}

\def\km{\hbox{km}}

\def\AU{\hbox{AU}}

\begin{document}

\title{Exploring the deep atmospheres of HD~209458b and WASP-43b using a non-gray general circulation model}
\titlerunning{Deep atmospheres of HD~209458b and WASP-43b}
\authorrunning{Schneider et al.}
\author{Aaron David~Schneider$^{1,2}$, Ludmila Carone$^{3}$, Leen Decin$^{2}$, Uffe Gr\r{a}e~J{\o}rgensen$^{1}$, Paul Molli\`{e}re$^{3}$, \\ Robin Baeyens$^{2}$, Sven Kiefer$^{2,4,5}$ \& Christiane Helling$^{4,5}$}
\institute{
  (1) Centre for ExoLife Sciences, Niels Bohr Institute, {\O}ster Voldgade 5, 1350 Copenhagen, Denmark\\
  (2) Institute of Astronomy, KU Leuven, Celestijnenlaan 200D, 3001, Leuven, Belgium\\
  (3) Max-Planck-Institut f\"ur Astronomie, K\"onigstuhl 17, 69117 Heidelberg, Germany\\
  (4) Space Research Institute, Austrian Academy of Sciences, Schmiedlstrasse 6, A-8042 Graz, Austria\\
  (5) TU Graz, Fakult\"at f\"ur Mathematik, Physik und Geod\"asie, Petersgasse 16, A-8010 Graz, Austria
} \date{\today}
\offprints{A. Schneider,\\ \email{aaron.schneider@nbi.ku.dk}\\or \email{aarondavid.schneider@kuleuven.be}}
\abstract{Simulations with a 3D general circulation model (GCM) suggest that one potential driver behind the observed radius inflation in hot Jupiters may be the downward advection of energy from the highly irradiated photosphere into the deeper layers. Here, we compare dynamical heat transport within the non-inflated hot Jupiter WASP-43b and the canonical inflated hot Jupiter HD~209458b, with similar effective temperatures. We investigate to what extent the radiatively driven heating and cooling in the photosphere (at pressures smaller than \SI{1}{\bar}) influence the deeper temperature profile (at pressures between \num{1} to \SI{700}{\bar}). Our simulations with the new non-gray 3D radiation-hydrodynamical model \texttt{expeRT/MITgcm} show that the deep temperature profile of WASP-43b is associated with a relatively cold adiabat. The deep layers of HD~209458b, however, do not converge and remain nearly unchanged regardless of whether a cold or a hot initial state is used. Furthermore, we show that different flow structures in the deep atmospheric layers arise. There, we find that WASP-43b exhibits a deep equatorial jet, driven by the relatively fast tidally locked rotation of this planet ($0.81$~days), as compared to HD~209458b ($3.47$~days). However, by comparing simulations with different rotation periods, we find that the resulting flow structures only marginally influence the temperature evolution in the deep atmosphere, which is almost completely dominated by radiative heating and cooling. Furthermore, we find that the evolution of deeper layers can influence the 3D temperature structure in the photosphere of WASP-43b. Thus, dayside emission spectra of WASP-43b may shed more light onto the dynamical processes occurring at greater depths.
}

\keywords{
Radiation: dynamics -- Radiative transfer -- Scattering -- Planets and satellites: atmospheres -- Planets and satellites: gaseous planets
}

\maketitle

\section{Introduction}\label{sec:introduction}
The \textit{James Webb Space Telescope} (JWST) is set to reveal data of unprecedented quality for several hot to ultra-hot Jupiters. Thanks to their inflated structure and hotter emission temperatures, these bodies are the easiest to observe, making them prime targets for further characterization. Furthermore, JWST observations may even yield insights into one of the long-standing mysteries in exoplanet science, which asks why many of the hot to ultra-hot Jupiters have large radii of $1.5 R_{\mathrm{Jup}}$ (inflated), while others have compact atmospheres (not inflated).

A number of processes have been proposed to explain hot Jupiter inflation,  including a reduction in the cooling efficiency or injection of additional energy into the interior. However, mechanisms that reduce the cooling of exoplanets fail to maintain inflated radii over timescales of billion years \citep[see the reviews of][]{Zhang2020review,Fortney2021review}. Recent studies have disfavored the reduced cooling hypothesis and point, rather, to the injection of stellar energy into the interior as a viable explanation. Examples include evidence of re-inflation in two close-in hot Jupiters as a result of their receiving more irradiation from their host stars during the course of their evolution \citep{Hartman2016Inflation} as well as a correlation between the amount of radius inflation and incident stellar flux \citep{Thorngren2018Inflation, Thorngren2019Interior}.

There are currently two favored processes for energy injection as an explanation for hot Jupiter inflation: a) Ohmic dissipation as initially proposed by \citet{Batygin2010Inflation}, where partially ionized wind flows are coupled with the planetary magnetic field, releasing heat in the interior; and b) dynamical downward energy transport from the irradiated atmosphere into deeper layers \citep{Tremblin2017Inflation,Sainsbury-Martinez2019Inflation,Sainsbury-Martinez2021Inflation}. In this study, we focus on the latter mechanism.

Understanding the vertical transport of stellar energy from irradiated atmospheres into deeper layers is computationally challenging as it requires investigating the temperature and the dynamical evolution of atmospheric layers below the observable photosphere (pressures much larger than $\SI{1}{\bar}$) with large runtimes of $\gg 10000$ simulation days \citep{Mayne2019GCM,Mendonca2019GCM,Wang2020GCM}. Recent studies have tried to tackle this computational issue by invoking additional assumptions at greater depths or by using a simplified radiative transfer treatment \citep[e.g.,][]{Tremblin2017Inflation, Mayne2017GCM, Mendonca2019GCM, Carone2020GCM,Sainsbury-Martinez2019Inflation,Sainsbury-Martinez2021Inflation}.

\citet{Mendonca2019GCM} performed a detailed analysis of angular momentum and heat transport using a Fourier decomposition of the angular momentum transport. The author found that the mean circulation deposits heat from the upper atmosphere into the deeper layers, where he concludes that additional energy sources are needed to explain the radius inflation. Using much longer integration times and a parametric radiative transfer scheme without heating in the deep atmosphere, \citet{Sainsbury-Martinez2019Inflation} instead found that radius inflation can be explained by dynamical heat transport if the deep atmosphere is not heated by irradiation. Both of these studies, along with that of  \citet{Showman2020review}, offer the conclusion that a self-consistent radiative transfer treatment is needed for the investigation of the deep layers to further quantify the coupling of the upper and lower atmospheres. Thus, there is a clear need for an efficient coupling mechanism between hydrodynamics and radiative transfer that is capable of monitoring deep processes over long simulation runtimes. We thus present \texttt{expeRT/MITgcm}\footnote{\texttt{expeRT/MITgcm} is an extension of the \texttt{exorad/MITgcm} scheme established in \citet{Carone2020GCM}.}, a fast and versatile radiative transfer scheme coupled with the 3D hydrodynamical code \texttt{MITgcm}.

The implementation of \texttt{expeRT/MITgcm} is described in Sect. \ref{sec:methods}. We apply the model to the exoplanets HD~209458b and WASP-43b in Section 3 and we explore the influence of layers deeper than \SI{1}{\bar} on the 3D thermal structure. HD~209458b is a well-observed and frequently modeled hot Jupiter, and WASP-43b is a JWST early release science target (ERS) \citep{Venot2020chem,Bean2018JWST}. More crucially, despite having similar effective temperatures between \SI{1400} - \SI{1500}{\K}, the first planet is inflated, whereas the latter is not \citep[e.g.,][Table 1 and Figure 1]{Helling2021CloudsParam}. We compare the behavior of the temperature convergence of WASP-43b and HD~209458b in the deep atmosphere in Sect. \ref{sec:temp_convergence}. We show the emission and transmission spectra for WASP-43b and HD~209458b in Sect. \ref{sec:spectra}. We discuss in Sect. \ref{sec:discussion} the need for GCMs that are capable of tackling different complex processes in hot Jupiters. Finally, we provide a summary and conclusion in Sect.~\ref{sec:conclusion}.

\section{Methods}\label{sec:methods}
In this section, we describe the implementation of radiative transfer into \texttt{expeRT/MITgcm}. We first describe the general circulation model in Sect. \ref{sec:dynamics} which is followed by a description of the radiative transfer extension introduced in this work (Sect. \ref{sec:RT}). The initialization of our simulations is discussed in Sect. \ref{sec:ic}.

\subsection{Dynamical core set-up}\label{sec:dynamics}
General circulation models (GCMs) are numerical frameworks that solve a (sub)set of the 3D equations of hydrodynamics on a rotating sphere. We integrated  \texttt{expeRT/MITgcm} with the deep wind GCM framework introduced in \citet{Carone2020GCM}, which uses the dynamical core of the \texttt{MITgcm} \citep{Adcroft2004}. The \texttt{MITgcm} uses an Arakawa C type cubedsphere grid with resolution C32\footnote{C32 is comparable to a resolution of 128 x 64 in longitude and latitude.} to solve the 3D hydrostatic primitive equations \citep[see e.g.,][Eqs. 1-5]{Showman2009GCM} assuming an ideal gas. We used the C32 grid with 24 cores in MPI multiprocessing, resulting in four 16x16 tiles for each of the six cubedsphere faces. As in \citet{Showman2009GCM}, we applied a fourth-order Shapiro filter with $\tau_{\mathrm{shap}}=\SI{25}{\s}$ to smooth horizontal grid-scale noise. The use of such smoothing schemes is common in GCMs, but may have a severe impact on atmospheric flows. We refer to \citet{Heng2011Benchmark, Skinner2021GCM} for a discussion on the impact of smoothing schemes. Our vertical grid uses 41 logarithmically spaced grid cells between \SI{1e-5}{\bar} and \SI{100}{\bar}. We extended the logarithmic domain by six linearly spaced grid cells between \SI{100}{\bar} and \SI{700}{\bar} in order to avoid a large spacing in the deep atmosphere.

We stabilized our model against nonphysical gravity waves by using a \enquote{soft} sponge layer that is similar to the one used in \texttt{THOR} \citep{Mendonca2018GCM, Deitrick2020GCM}. The zonal horizontal velocity $u$ $[\SI{}{\m\per\s}]$ is damped by a Rayleigh friction term towards its longitudinal mean $\bar u$ $[\SI{}{\m\per\s}]$ via:
\begin{equation}\label{eq:sponge}
        \frac{\mathrm{d}u}{\mathrm{d}t} = - \tilde k_\mathrm{top} (u - \bar u),
\end{equation}
where $t$ \SI{}{[\s]} is the time and $\tilde k_\mathrm{top}$ \SI{}{[\s^{-1}]} is the strength of the opposed friction and is calculated as a function of pressure $p$ \SI{}{[\Pa]} by
\begin{equation}
        \tilde k_\mathrm{top}(p) = k_\mathrm{top} \cdot \max\left[0, 1-\left(\frac{p}{p_\mathrm{sponge}}\right)^2 \right]^2.
\end{equation}
The fudge parameters $p_\mathrm{sponge}$ \SI{}{[\Pa]} and $k_\mathrm{top}$ \SI{}{[\s^{-1}]} are used to control the position and the strength of the opposed Rayleigh friction. Throughout this paper, we use $k_\mathrm{top}=\SI{20}{\per\day}$ and $p_\mathrm{sponge}=\SI{1e-4}{\bar,}$ which results in a friction term that slowly increases in strength for $p\leq\SI{1e-4}{\bar}$. We want to note here that we calculate Eq.~\ref{eq:sponge} by deprojecting the cubed sphere grid onto its geographic direction, averaging the deprojected $u$ over 20 latitudinal bins and projecting the resulting $\bar u$ on the cubedsphere grid.

To parametrize the deep magnetic drag and to stabilize the lower boundary, we followed \citet{Carone2020GCM}. We imposed additional Rayleigh friction to the winds in the deep layers ($p \geq \SI{490}{\bar}$). Within these layers we damp the zonal velocity $u$ $[\SI{}{\m\per\s}]$ and meridional velocity $v$ $[\SI{}{\m\per\s}]$ by
\begin{equation}\label{eq:deep}
        \frac{\mathrm{d}u}{\mathrm{d}t} = - \frac{u}{\tilde\tau_\mathrm{deep}},~~
        \frac{\mathrm{d}v}{\mathrm{d}t} = - \frac{v}{\tilde\tau_\mathrm{deep}},
\end{equation}
where we set the strength of the drag $\tilde\tau_\mathrm{deep}$ $[\SI{}{\s}]$ as
\begin{equation}
        \frac{1}{\tilde\tau_\mathrm{deep}} = \frac{1}{\tau_\mathrm{deep}}~\max \left(0,\frac{p-\SI{490}{\bar}}{\SI{700}{\bar}-\SI{490}{\bar}}\right).
\end{equation}
We used the same parameters as in \citet{Carone2020GCM} and set $\tau_\mathrm{deep} = \SI{1}{\day}$.\footnote{There is a typo in the drag description in \citet{Carone2020GCM}: $\tau_\mathrm{fric}$ and $\tau_\mathrm{bottom,fric}$ should be $\frac{1}{\tau_\mathrm{fric}}$ and $\frac{1}{\tau_\mathrm{bottom,fric}}$ respectively.}
\citet{Carone2020GCM} showed that these measures are important for maintaining numerical stability and to prevent nonphysical boundary effects for WASP-43b. These authors found that models with and without the deep drag yield small differences and only in the temperature and dynamics.

We did not use a convective adjustment scheme to model convection on a subgrid scale \citep[e.g.,][]{Deitrick2020GCM, Lee2021GCM} because we are explicitly interested in the formation and stability of a deep adiabat by vertical heat-transport. \citet{Sainsbury-Martinez2019Inflation} argued that such a deep adiabat should form due to the absence of radiative heating and cooling at layers higher up in the atmosphere compared to layers where the atmosphere would become unstable to convection according to the Schwarzschild criterion.

The gas temperature $T$ \SI{}{[\K]} of an ideal gas relates the density with the pressure and therefore ultimately determines the dynamics. \texttt{MITgcm} treats the temperature by means of the potential temperature, $\Theta$ \SI{}{[\K],} which is given by
\begin{equation}\label{eq:Theta}
        \Theta = T \left(\frac{p_0}{p}\right)^{R/c_p},
\end{equation}
where $c_p$ $[\SI{}{\J\per\kg\per\K}]$ is the heat capacity at constant pressure, $p_0$ $[\SI{}{Pa}]$ is the pressure at the bottom of the computational domain, and $R$ $[\SI{}{\J\per\kg\per\K}]$ is the specific gas constant.
The potential temperature $\Theta$ is then forced using the thermodynamic energy equation \citep[e.g.,][]{Showman2009GCM}:
\begin{equation}\label{eq:energyeq}
        \frac{\mathrm{d}\Theta}{\mathrm{d}t} = \frac{\Theta}{T}\frac{q}{c_p},
\end{equation}
where $q$ $[\SI{}{\W\per\kg}]$ is the heating rate. The heating rate $q$ is given by
\begin{equation}\label{eq:hr}
        q = g \frac{\partial F^\mathrm{net}}{\partial p},
\end{equation}
where $g$ $[\SI{}{\m\per\s\squared}]$ is the gravity and $F^\mathrm{net}$ $[\SI{}{\W\per\m\squared}]$ is the net radiative flux. We note that this approach towards radiative heating and cooling is not necessarily flux conserving, yielding the possibility of net cooling or heating of the planet.

\subsection{Radiative transfer}\label{sec:RT}
We used the Feautrier method \citep{Feautrier1964} in combination with lambda iteration in order to solve the radiative transfer equation. These schemes are frequently used in works on stellar atmospheres \citep[e.g.,][]{Gustafsson1975c-k, Gustafsson2008MARCS}, protoplanetary disks \citep{Dullemond2002pp} and have been applied in some 1D models of exoplanetary atmospheres \citep{Molliere20151Dmodel, Molliere20171Dmodel, Molliere20191Dmodel, Molliere20201Dmodel,Piette2020RT}. Our implementation of radiative transfer adopts the radiative transfer scheme used in the 1D planet atmosphere model \texttt{petitRADTRANS} \citep{Molliere20191Dmodel, Molliere20201Dmodel}.

\subsubsection{Fluxes}
\label{sec:RT_flux}
The change of intensity $I_\nu$ $[\SI{}{\W\per\m\squared\per\steradian\per\Hz}]$ along a ray passing through a planetary atmosphere in the direction of the unit vector, $\mathbf{n,}$ may be described by the radiative transfer equation
\begin{equation}\label{eq:radtrans}
        \mathbf{n}\cdot \nabla I_\nu = \alpha^\mathrm{tot}_\nu \left(S_\nu - I_\nu \right),
\end{equation}
where $\nu$ $[\SI{}{\Hz}]$ is the frequency, $S_\nu$ $[\SI{}{\W\per\m\squared\per\steradian\per\Hz}]$ is the source function and $\alpha^\mathrm{tot}_\nu$ $[\SI{}{\m^{-1}}]$ is the inverse mean-free path of the light ray. We solve this equation for photons originating within the planetary atmosphere or scattered out of the incoming stellar ray. The extinction of the incoming stellar intensity is separately modeled using an exponential decay. This is not an approximation; such a separate treatment is allowed given to the linear nature of the equation of radiative transfer. The full solution is then obtained from adding the two intensities (planetary and scattered stellar photons along with$ $ extincted stellar intensity). The inverse mean free path $\alpha^\mathrm{tot}_\nu $ is given by the sum of absorption and scattering inverse mean free paths and may be written as
\begin{equation}
        \alpha^\mathrm{tot}_\nu = \alpha^\mathrm{abs}_\nu +\alpha^\mathrm{scat}_\nu.
\end{equation}
The mean intensity $J_\nu$ $[\SI{}{\W\per\m\squared\per\steradian\per\Hz}]$ is the zeroth order radiative momentum and is given by angle integration (e.g., Gaussian quadrature) of the intensity field. The zeroth and first order radiative moments in plane-parallel are given by
\begin{equation}\label{eq:rad_moments}
        \left[J_\nu, H_\nu \right] = \frac{1}{2}\int_{-1}^{1}\left[1,\mu\right] I_\nu(z,\mu) \mathrm{d}\mu,
\end{equation}
where $\mu=\cos\theta$ is the angle between the atmospheric normal and the ray and $H_\nu$ $[\SI{}{\W\per\m\squared\per\Hz}]$ is related to the flux (see below).

We use the source function in the coherent isotropic scattering approximation for the planetary and scattered stellar photon field, given by
\begin{equation}\label{eq:source}
        S_\nu = \epsilon_\nu B_\nu(T) + (1-\epsilon_\nu) (J_\nu^\mathrm{pla} + J_{\nu}^\star),
\end{equation}
where $B_\nu(T)$ $[\SI{}{\W\per\m\squared\per\steradian\per\Hz}]$ is the Planck function at temperature $T$ $[\SI{}{\K}]$, while $J_\nu^\mathrm{pla}$ $[\SI{}{\W\per\m\squared\per\steradian\per\Hz}]$ and $J_{\nu}^\star$ $[\SI{}{\W\per\m\squared\per\steradian\per\Hz}]$ are the mean intensity of the planetary atmosphere and the stellar attenuated radiation field respectively, and $\epsilon_\nu$ is the photon destruction probability given by
\begin{equation}\label{eq:photon_destruct}
        \epsilon_\nu = \frac{\alpha^\mathrm{abs}_\nu}{\alpha^\mathrm{tot}_\nu}.
\end{equation}
The mean intensity of the stellar attenuated radiation field in a plane-parallel atmosphere $J^\star_{\nu}$ $[\SI{}{\W\per\m\squared\per\steradian\per\Hz}]$ \citep{Molliere20151Dmodel} is given by
\begin{equation}\label{eq:J_ini}
        J^\star_{\nu} = \frac{I^\star_{\nu}(p=0)}{4\pi}\exp\left(-\tau_\nu/\mu_*\right),
\end{equation}
where $I^\star_\nu(p=0)$ $[\SI{}{\W\per\m\squared\per\steradian\per\Hz}]$ is the stellar intensity field at the top of the planetary atmosphere. Here, \texttt{expeRT/MITgcm} uses the PHOENIX stellar model spectrum \citep{Husser2013PHOENIX} that is part of \texttt{petitRADTRANS} to calculate the stellar attenuated radiation field. The angle between the atmospheric normal and the incoming stellar light $\mu_\star$ for a tidally locked exoplanet is given by
\begin{equation}
    \label{eq:mu_star_definition}
        \mu_\star = \cos\vartheta\cos\varphi,
\end{equation}
where $\vartheta$ is the latitude and $\varphi$ is the longitude. We note that we do not extend the plane-parallel assumption by including a geometric depth dependence in $\mu_\star$, which would improve the treatment of the poles (see Appendix~\ref{sec:planeparallel}).
The optical depth, $\tau_\nu,$ in a hydrostatic atmosphere that is parallel to the atmospheric normal is given by
\begin{equation}\label{eq:tau}
        \tau_\nu = \int \frac{\kappa_\nu^\mathrm{tot}}{g} \mathrm{d}p,
\end{equation}
where $\kappa_\nu^\mathrm{tot}$ $[\SI{}{\m\squared\per\kg}]$ is the cross-section per unit mass (opacity) and depends on the gas density, $\rho$ $[\SI{}{\kg\per\m\cubed}],$ as:
\begin{equation}
        \kappa_\nu^\mathrm{tot} = \frac{\alpha^\mathrm{tot}_\nu}{\rho}.
\end{equation}
We can express the received bolometric stellar flux $F^\star$ $[\SI{}{\W\per\m\squared}]$ using Eqs.~\ref{eq:rad_moments} and \ref{eq:J_ini} as
\begin{equation}\label{eq:F_star}
        F^\star = -4 \pi \int_\nu H^\star_{\nu}\,\mathrm{d}\nu
=  4\pi\mu_\star \int_\nu J^\star_{\nu}\, \mathrm{d}\nu.
\end{equation}
We solved Eq.~\ref{eq:radtrans} using the well established Feautrier method \citep{Feautrier1964} with accelerated lambda iteration \citep[ALI,][]{Olson1986} in order to get the planetary flux. Once a converged solution (see Sect. \ref{sec:source_convergence}) of the planetary intensity field has been found, we can calculate the emerging bolometric fluxes from the integration of the planetary radiation field $F^\mathrm{pla}$ $[\SI{}{\W\per\m\squared}]$ using Eq. \ref{eq:rad_moments}:
\begin{equation}
        F^\mathrm{pla} = \int_\nu F^\mathrm{pla}_\nu \mathrm{d}\nu = -4 \pi \int_\nu H^\mathrm{pla}_\nu\mathrm{d}\nu.
\end{equation}

The total bolometric flux $F^\mathrm{net}$ $[\SI{}{\W\per\m\squared}]$ can now be calculated by
\begin{equation}\label{eq:netflux}
        F^\mathrm{net} = F^\mathrm{pla} + F^\star.
\end{equation}

Since Eq.~\ref{eq:hr} requires finite differencing of $F^\mathrm{net}$, we followed the approach of \citet{Showman2009GCM} and evaluated the fluxes on the vertically staggered cell interfaces. \citet{Lee2022BDWD} find that the use of a quadratic Bezier interpolation scheme for the interpolation of the temperature to the vertical interfaces greatly improves the accuracy and stability of the radiative transfer routine. Our tests confirm this finding and we therefore chose to also use this interpolation scheme.

\begin{figure}
        \centering
        \includegraphics[width =.45\textwidth]{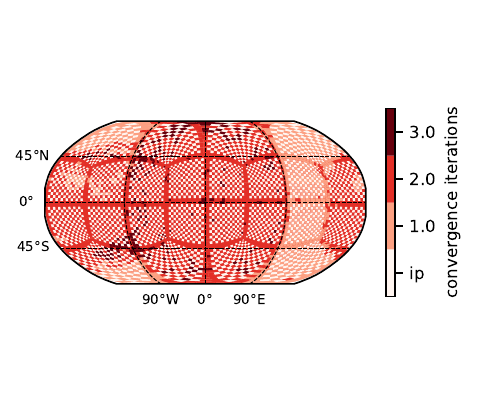}
        \caption{Horizontal map displaying the number of iterations to converge the source function (Eq. \ref{eq:source}) during runtime (at $t=\SI{12000}{\day}$) for the HD~209458b simulation. The interpolated cells (ip) are staggered to each other. The C32 grid divides the horizontal domain into 24 individual computational domains, as seen by the line structure.}
        \label{fig:convergence}
\end{figure}

We calculated Eq.~\ref{eq:netflux} for every second grid column and linearly interpolate the fluxes in between \citep[see e.g.,][]{Showman2009GCM}. The placement of the interpolated cells is shown in Fig. \ref{fig:convergence}, where the bright cells indicate the interpolated cells. We note that the C32 geometry prohibits a strict interpolation of every second column. This is caused by the tiling of the horizontal domain into individual computational domains. It is therefore necessary to calculate the radiative transfer at the tile borders, as seen in Fig. \ref{fig:convergence}. We test the interpolation scheme in Appendix~\ref{sec:interp_vs_nointerp} and find that the error introduced by the interpolation is below 2\%.

\subsubsection{Convergence of the source function}\label{sec:source_convergence}

%
%
%

There are two modes that \texttt{expeRT/MITgcm} can operate in: with scattering and without scattering. We set $\epsilon_\nu=1$ if scattering is neglected, in which case Eq.~\ref{eq:source} can be reduced to the LTE source function $S_\nu = B_\nu(T)$.
We iterate over the radiative transfer solution (so-called lambda iteration) if $\epsilon_\nu<1$ (i.e., scattering is included), since $S_\nu$ depends on $I_\nu$ (see Eqs.~\ref{eq:rad_moments} and \ref{eq:source}). We note that the Feautrier method has an exact solution in the isotropic scattering case. However, its solution would require a full matrix inversion \citep[see e.g.,][]{Hubeny2017RT}, which is computationally expensive. Here we chose to solve the Feautrier equation by inverting a tridiagonal matrix instead, which requires iterating on the source function. For this, we use accelerated lambda iteration instead \citep[ALI,][]{Olson1986} to speed up the convergence of the source function. We choose to call the source function converged if the relative change of the source function is smaller than 2\% of the reciprocal local lambda approximator \citep{Olson1986,Rutten2003}. We performed numerical experiments, which confirmed that 2\% yields a good tradeoff between accuracy and model performance, in agreement with \citet{Auer1991RT,Rutten2003}. We use the knowledge of the source function of the previous radiative time-step as an initial guess for the current time-step. The number of iterations that are needed to converge the source function depends on the amplitude of structural changes in the temperature with time. During the first radiative time-step (see Table~\ref{tab:radtimesteps}) we find that 50-100 iterations are needed to converge the source function, while later time-steps reduce the amount of necessary iterations significantly to orders of one to three iterations per time-step (see Fig.~\ref{fig:convergence}).

Our numerical implementation of the ALI method in \texttt{expeRT/MITgcm} is an optimized version of the flux routine (with scattering) in \texttt{petitRADTRANS} \citep{Molliere20201Dmodel}, which has been tested and benchmarked \citep{Baudino2017}. The major difference between the implementation in \texttt{petitRADTRANS} and \texttt{expeRT/MITgcm} is that we chose to not include Ng acceleration \citep{Ng1974} since the inclusion of Ng acceleration would only help within the first few time steps, whereas convergence of the source function is reached within a few iterations at later time steps (usually less than 2, see also Fig. \ref{fig:convergence}).

The flexibility of our radiative transfer approach combined with its accuracy yields no drawback in terms of performance. When using the 24 core model setup, we report fast model runtimes of approximately 600 or 1000 simulated days per day of model runtime for a radiative time-step of $\SI{100}{\s}$ or $\SI{300}{\s}$, respectively. We note that we take a different approach than e.g., \citet{Rauscher2012double, Lee2021GCM}, who developed a less complex radiative treatment to shorten the model run-time. Instead, our approach uses a full radiative transfer treatment, but with a different implementation than the one used in previous GCMs \citep[e.g.,][]{Showman2009GCM, Amundsen2016UKMetGCM}.

\subsubsection{Opacities}
\begin{table}
        \caption{Opacity table}
        \centering
                \begin{tabular}{c c c}
                \hline\hline
                Type & Species & source\\
                \hline
                Gas & H$_2$O & \citet{Polyansky2018Opac}\\
                & CO$_2$ & \citet{Yurchenko2020Opac}\\
                & CH$_4$ & \citet{Yurchenko2017Opac}\\
                & NH$_3$ & \citet{Coles2019Opac}\\
                & CO & \citet{Li2015Opac}\\
                & H$_2$S & \citet{Azzam2016Opac}\\
                & HCN & \citet{Barber2014Opac}\\
                & PH$_3$ & \citet{Sousa-Silva2015Opac}\\
                & TiO & \citet{McKemmish2019Opac}\\
                & VO & \citet{McKemmish2016Opac}\\
                & FeH & \citet{Wende2010Opac}\\
                & Na & \citet{Piskunov1995Opac}\\
                & K & \citet{Piskunov1995Opac}\\
                \hline
                Rayleigh & H$_2$ & \citet{Dalgarno1962Opac}\\
                scattering& He & \citet{Chan1965Opac}\\
                \hline
                CIA & H$_2$-H$_2$ & BR, RG12\\
                & H$_2$-He & BR, RG12\\
                & H$^-$ & \citet{Gray2008Opac}\\
                \hline
        \end{tabular}
        \begin{tablenotes}
                 \item \textbf{Notes:} CIA is short for collision induced absorption. BR stands for \citet{Borysow1988Opac, Borysow1989aOpac, Borysow1989bOpac, Borysow2001Opac, Borysow2002Opac} and RG12 stands for \citet{Richard2012Opac}. Potassium and sodium are broadened with the coefficients of \citet{Allard2019OpacNa}.
        \end{tablenotes}
        \label{tab:opac}
\end{table}

\begin{table}
        \centering
        \caption{Low-resolution wavelength grid (S0)}
        \begin{tabular}{c c}
                \hline\hline
                Left border wavelength [\SI{}{\mu\m}] & Right border wavelength [\SI{}{\mu\m}]
                \\\hline
                0.26 & 0.42\\
                0.42 & 0.85\\
                0.85 & 2.02\\
                2.02 & 3.50\\
                3.50 & 8.70\\
                8.70 & 300.00\\
                \hline
        \end{tabular}
        \begin{tablenotes}
                \item \textbf{Notes:} Low-resolution wavelength grid (S0) used throughout this work. Rows represent one wavelength bin.
        \end{tablenotes}
        \label{tab:opac_bins}
\end{table}

\begin{table}
        \centering
        \caption{Wavelength resolutions}
        \begin{tabular}{c c c}
                \hline\hline
                Abbreviation & $N_\mathrm{bins}$ & Reference
                \\\hline
                S0 & 6 & Table \ref{tab:opac_bins}\\
                S1 & 11 & \citet{Kataria2013}\\
                S2 & 30 & \citet{Showman2009GCM}\\
                \hline
        \end{tabular}
        \begin{tablenotes}
                \item \textbf{Notes:} Wavelength resolutions used in \texttt{expeRT/MITgcm}. We note that the upper wavelength edge of all methods (including S1 and S2) is set to \SI{300}{\mu\m}. The accuracy of these different binning methods has been benchmarked in Appendix \ref{sec:tests}.
        \end{tablenotes}
        \label{tab:opac_bins_overview}
\end{table}

Generally, \texttt{expeRT/MITgcm} operates on a precalculated grid of correlated-k opacities. The opacity in each correlated-k \citep[e.g.,][]{Goody1989c-k} wavelength bin is sorted in size and its distribution is sampled by a 16 point subgrid. Opacities are precalculated offline on a grid of 1000 logarithmically spaced temperature points between \SI{100}{K} and \SI{4000}{K} for every vertical layer. An overview of the used opacity sources can be found in Table \ref{tab:opac}. Opacities were obtained from the \texttt{ExoMolOP} database \citep{Chubb2021Opac} where available.

Opacities are precalculated and combined using a wrapper around the low-resolution mode of \texttt{petitRADTRANS}. Molecular abundances, the value for the specific gas constant $R=\SI{3590}{\J\per\kg\per\K}$ and the specific heat capacity at constant pressure, $c_p=\SI{12766}{\J\per\kg\per\K}$, are inferred using the equilibrium chemistry package of \texttt{petitRADTRANS}. We used solar metallicity and solar C/O ratios, however, formation models for hot gas giants may hint towards sub-solar C/O ratios and super-solar metallicities \citep[e.g.,][]{Mordasini2016,Schneider2021I,Schneider2021II}. We note that our approach to the chemical composition of the gas is computationally fast, but implies a fixed chemical composition during runtime. Additionally, this treatment neglects a physically correct treatment of cold traps, where certain species would condense out and not be available at altitudes above, which could prohibit a stratosphere caused by TiO and VO.

We used \texttt{expeRT/MITgcm} with three different wavelength grids (see Tables \ref{tab:opac_bins} and \ref{tab:opac_bins_overview}). The default resolution S0 uses $N_\mathrm{bins}=\num{6}$ wavelength bins from \SI{0.26}{\mu\m} to \SI{300}{\mu\m}. We used finer grids with \num{11} or \num{30} bins (S1 and S2 hereafter) to benchmark our model to the wavelength grids used in \texttt{SPARC/MITgcm} \citep{Kataria2013}. Those tests are shown in Appendix \ref{sec:tests}.

\subsection{Model initialization}\label{sec:ic}
We followed the suggestion of \citet{Lee2021GCM} and homogeneously initialize the upper layers (pressures below \SI{1}{\bar}) of our models with the analytic temperature profile of \citet{Parmentier20151D}. We calculated the substellar irradiation temperature $T_\mathrm{irr}$ of HD~209458b and WASP-43b using the stellar radius, $R_\star$ $[\SI{}{\m}]$, semi major axis, $a_p$ $[\SI{}{\m}],$ and stellar effective temperature, $T_\star$ $[\SI{}{\K}],$ as
\begin{equation}\label{eq:t_eff}
        T_\mathrm{irr} = T_\star \sqrt{\frac{R_\star}{2a_p}},
\end{equation}
where we use an effective stellar temperature, $T_\star$, and a stellar radius, $R_\star$, of $T_\star^{\mathrm{HD~209458}}=\SI{6092}{\K}$ and $R_\star^{\mathrm{HD~209458}}=\SI{1.203}{R_\odot}$(\citet{Boyajian2015HD2}) for HD~209458b and then  $T_\star^{\mathrm{WASP-43}}=\SI{4520}{\K}$ and $R_\star^{\mathrm{WASP-43}}=\SI{0.667}{R_\odot}$ for WASP-43b \citep{Gillon2012Obs}. We assumed both planets to be tidally locked on a circular orbit. We set the rotation period of HD~209458b to $P_\mathrm{rot}^{\mathrm{HD~209458b}}=\SI{3.47}{\day}$ and the orbital separation to $a_p^{\mathrm{HD~209458b}}=\SI{0.04747}{\AU}$ \citep{Southworth2008HD2,Southworth2010HD2} and for WASP-43b to $P_\mathrm{rot}^{\mathrm{WASP-43b}}=\SI{0.8135}{\day}$ and $a_p^{\mathrm{WASP-43b}}=\SI{0.01526}{\AU}$ \citep{Gillon2012Obs}.
Using Eq.~\ref{eq:t_eff}, we estimated $T_\mathrm{irr}^{\mathrm{HD~209458b}}=\SI{1479}{\K}$ and $T_\mathrm{irr}^{\mathrm{WASP-43b}}=\SI{1441}{\K}$ for HD~209458b and WASP-43b, respectively. We used the fit from \citet{Thorngren2019Interior} with these irradiation temperatures to estimate the interior temperature resulting from radiation of the planetary interior. This procedure yields interior temperatures of \SI{575}{\K} and \SI{549}{\K}, respectively. These irradiation and interior temperatures are then used to calculate the analytic temperature profile of \citet{Parmentier20151D}. We note that these interior temperatures were only used to initialize the temperature of our simulations, since we do not enforce radiative (-convective) equilibrium in the atmosphere of our GCM models during runtime.

The deep layers (pressures above \SI{10}{\bar}) are homogeneously set to a hot adiabatic temperature profile of the form
\begin{equation}
        T(p>\SI{10}{\bar}) = \Theta_\mathrm{ad} \cdot \left(\frac{p}{\SI{1}{\bar}}\right)^\eta,
\end{equation}
where we follow \citet{Sainsbury-Martinez2019Inflation} to guess high values of $\Theta_\mathrm{ad}^{\mathrm{HD~209458b}} = \SI{1800}{\K}$ and $\Theta_\mathrm{ad}^{\mathrm{WASP-43b}} = \SI{1400}{\K}$ as the temperature of the adiabat at \SI{1}{\bar}. The slope of the adiabat $\eta$ is given by
\begin{equation}
        \eta = \frac{R}{c_p}\approx 3.56.
\end{equation}

The intermediate pressure ($\SI{1}{\bar}\leq p \leq \SI{10}{\bar}$) levels are linearly interpolated between the adiabat and the analytic model of \citet{Parmentier20151D}. This procedure allows for a hot initial state in the deep layers, as proposed in \citet{Sainsbury-Martinez2019Inflation}. We assume a value of $g^{\mathrm{HD~209458b}}=\SI{8.98}{\m\per\s\squared}$ \citep[e.g.,][]{Lee2021GCM} and $g^{\mathrm{WASP-43b}}=\SI{46.9}{\m\per\s\squared}$ \citep{Gillon2012Obs} for the surface gravity, respectively. Differences in the model setup of WASP-43b compared to HD~209458b (see Table \ref{tab:param}) only occur in the surface gravity, rotation period and radiative time-step (see Table \ref{tab:radtimesteps}).

We ran the model for \num{12000} days with a dynamical time-step of $\Delta t = \SI{25}{\s}$.\footnote{Day refers to \SI{24}{\hour} or 1 earth day.} Since the temperature in the upper layers might change significantly during the first 500 days, we recalculate the radiative fluxes in every fourth dynamical time-step, which we do by defining a radiative time-step, $\Delta t_\mathrm{rad}$, which we set to $\Delta t_\mathrm{rad}=\SI{100}{\s}$ (see Table \ref{tab:radtimesteps}). In the case of WASP-43b, we continue the simulation with the same radiative time-step, whereas we use a longer radiative time-step for the rest of the HD~209458b simulation of \SI{300}{\s}. A shorter radiative time-step is needed for the simulations of WASP-43b, since WASP-43b undergoes large structural changes throughout the model runtime, whereas HD~209458b reaches a steady state in the observable atmosphere within the first few hundred days (see Sect. \ref{sec:temp_convergence}).

\section{State of the upper atmosphere}\label{sec:results}
The number of planets modelled with fully coupled GCMs has increased a great deal since \citet{Showman2009GCM} first introduced a fully coupled GCM for hot Jupiters. We chose to focus on HD~209458b and WASP-43b in this work. HD~209458b has been modeled by many of the relevant fully coupled hot Jupiter GCMs \citep{Showman2009GCM, Amundsen2016UKMetGCM, Lee2021GCM}, making it a suitable target for comparisons between our model and previous works. Our non-gray 3D GCM WASP-43b simulations will be compared to simulations, at a similar level of complexity, of the same planet by \citet{Kataria2015WASP}. Comparisons with \cite{Baeyens2021GCM}, where a simpler radiative transfer scheme is used, are shown in Appendix \ref{sec:newton}.

\subsection{HD~209458b}\label{sec:HD2}
\begin{figure*}
        \centering
        \includegraphics[width=\textwidth]{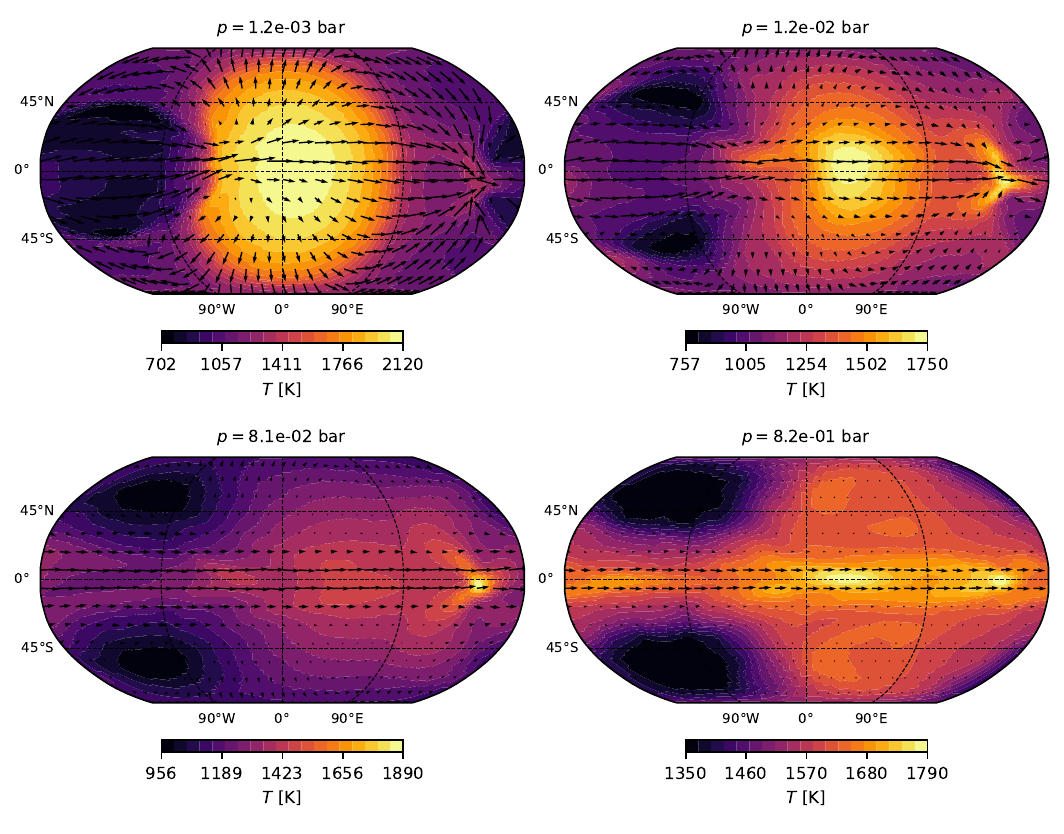}
        \caption{Maps of HD~209458b displaying the color coded temperature at different pressure levels (as indicated above the maps). The substellar point is located at (0$^\circ$,0$^\circ$).}
        \label{fig:HD2_temp}
\end{figure*}

\begin{figure}
        \centering
        \includegraphics[width=.45\textwidth]{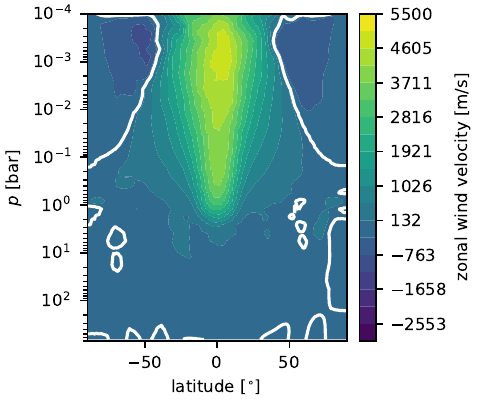}
        \caption{Zonally averaged zonal wind velocities for HD~209458b. Positive values correspond to prograde flow. The white line depicts the change of sign.}
        \label{fig:HD2_zonal_mean}
\end{figure}
We show the time-averaged temperature for different pressure layers in Fig. \ref{fig:HD2_temp}. The temperature has been time-averaged over the last \SI{100}{\day}. The temperature maps demonstrate the typical features of a tidally locked hot Jupiter climate: The strongly irradiated dayside transports heat to the nightside via a superrotating jet \citep[e.g.,][]{Showman2011Superrotation}, while shifting the hottest point eastwards. The upper layers ($p<\SI{1e-3}{\bar}$) are strongly influenced by stellar irradiation, whereas advection is negligible (Fig.~\ref{fig:HD2_temp}, top left panel). In contrast, the deeper layers ($p>\SI{1e-2}{\bar}$) are more efficient in redistributing heat by advection (Fig.~\ref{fig:HD2_temp}, bottom left panel).

The zonal mean wind velocities are shown in Fig. \ref{fig:HD2_zonal_mean}. The superrotating jet of HD~209458b as seen in \citet{Kataria2016GCM,Amundsen2016UKMetGCM,Lee2021GCM} is reproduced with \texttt{expeRT/MITgcm}. The magnitude of the jet ($\approx\SI{5000}{\m\per\s}$) agrees best with \citet{Kataria2016GCM, Lee2021GCM} and is lower than the value reported by \citet{Amundsen2016UKMetGCM} of \SI{7000}{\m\per\s}.

The temperature maps of the HD~209458b simulations (Fig. \ref{fig:HD2_temp}) qualitatively match those of \citet[][Fig.~5]{Lee2021GCM} showing that the general flow pattern as well as the day to night temperature difference are similar. Small asymmetric features in the temperature map highlight that the model has not yet equillibriated, even though the model was run for 12000 days. We note that our simulations are initialized with a hot deep adiabat that is maintained throughout the runtime (see Sect. \ref{sec:temp_convergence}) while other simulations of HD~209458b \citep[e.g.,][]{Amundsen2016UKMetGCM, Showman2009GCM} did not use such a hot interior.

The simulations of \citet{Showman2009GCM, Lee2021GCM} exhibit a temperature inversion due to the presence of TiO and VO around the substellar point, which results in an enhanced absorption of incoming stellar light. Our simulation of HD~209458b reproduces this temperature inversion, since we also include contributions of TiO and VO to the gas opacities that we use in our models (see Table~\ref{tab:opac}). Small uncertainties in temperature and opacities may result in slightly hotter or cooler gas temperatures, which could result in the presence or absence of TiO and VO, since the dayside temperature of HD~209458b is very close to the TiO and VO condensation curves. This further highlights the importance of accurate interpolation schemes, such as the Bezier interpolation for the temperature (see Sect.~\ref{sec:RT_flux}), as well as the fine grained opacity grid with 1000 logarithmically spaced temperature points.

\subsection{WASP-43b}\label{sec:WASP-43b}
\begin{figure*}
        \centering
        \includegraphics[width=\textwidth]{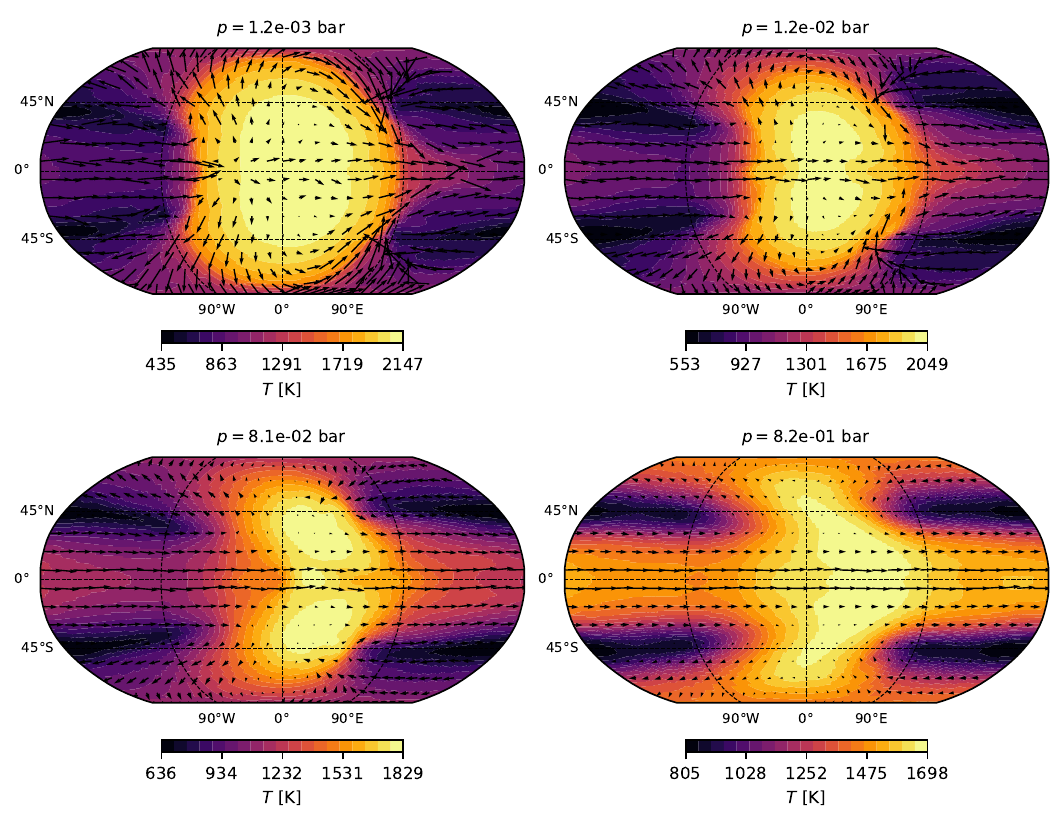}
        \caption{Maps of WASP-43b displaying the color coded temperature at different pressure levels (like Fig.~\ref{fig:HD2_temp}).}
        \label{fig:WASP-43b_temp}
\end{figure*}
\begin{figure}
        \centering
        \includegraphics[width=.45\textwidth]{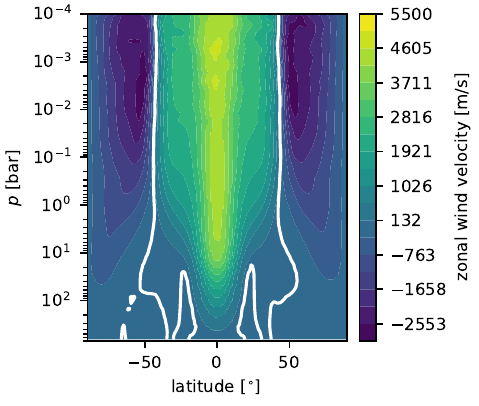}
        \caption{Zonally averaged zonal wind velocities (like Fig. \ref{fig:HD2_zonal_mean}) for WASP-43b.}
        \label{fig:WASP-43b_zonal_mean}
\end{figure}

Similarly to Fig.~\ref{fig:HD2_temp} for HD~209458b, the temperature maps for WASP-43b are shown in Fig.~\ref{fig:WASP-43b_temp}. The results are well in line with the \texttt{SPARC/MITgcm} results for WASP-43b \citep{Kataria2015WASP} with TiO and VO. The day-to-night temperature differences as well as the flow patterns are comparable.

Unlike the case of \citet{Carone2020GCM}, our simulations do not exhibit a retrograde flow. As noted in \citet{Carone2020GCM}, the retrograde flow in their simulation was embedded in strong superrotation. Thus, both tendencies - the one of the prograde flow and that of the retrograde flow - were engaged in a \enquote{tug of war}. The main difference between the simulations of \citet{Carone2020GCM} and this work is, however, that this work does not require artificial temperature forcing in deep layers. Our simulations therefore capture the full feedback between dynamics and radiation that was missing in \citet{Carone2020GCM}. This feedback seems to strengthen superrotation, apparently allowing for full superrotation to win the \enquote{tug of war} over the tendency for retrograde flow. \citet{Carone2020GCM} found that the tendency toward retrograde flow is associated with a deep jet that can transport zonal momentum upwards. The transport of zonal momentum has been analyzed in several studies \citep[e.g.,][]{Showman2015GCM,Mendonca2019GCM, Wang2020GCM, Carone2020GCM}, but it is still not clear what conditions would lead a deep equatorial jet toward retrograde flow. We will investigate the transport of zonal momentum in more detail in a follow-up publication and instead focus on the temperature evolution at greater depths in this work.

\citet{Carone2020GCM} find that the depth of the equatorial jet of WASP-43b is likely linked to the rotation period. Their argumentation bases on two simulations of WASP-43b. Their nominal WASP-43b simulation with a rotation period of \num{0.81} days shows signs of a deep equatorial jet, which is not found in a simulation of WASP-43b with a rotation period of \num{3.5} days (as for HD~209458b). Our non-gray model of WASP-43b confirms this finding of \citet{Carone2020GCM}. WASP-43b indeed exhibits a deeper equatorial jet compared to HD~209458b, which can be seen in the zonal mean of the wind flow (see Fig. \ref{fig:WASP-43b_zonal_mean}). We note that the simulations of \citet{Kataria2015WASP} do not exhibit such a deep jet, which is likely caused by the short runtime of \SI{50}{\day} compared to the \SI{12000}{\day} used here. Future works are warranted to unveil the physical reasons behind the depth of the equatorial jet.

\section{Temperature convergence of the deep layers}\label{sec:temp_convergence}

\begin{figure*}
        \centering
        \includegraphics[width=\textwidth]{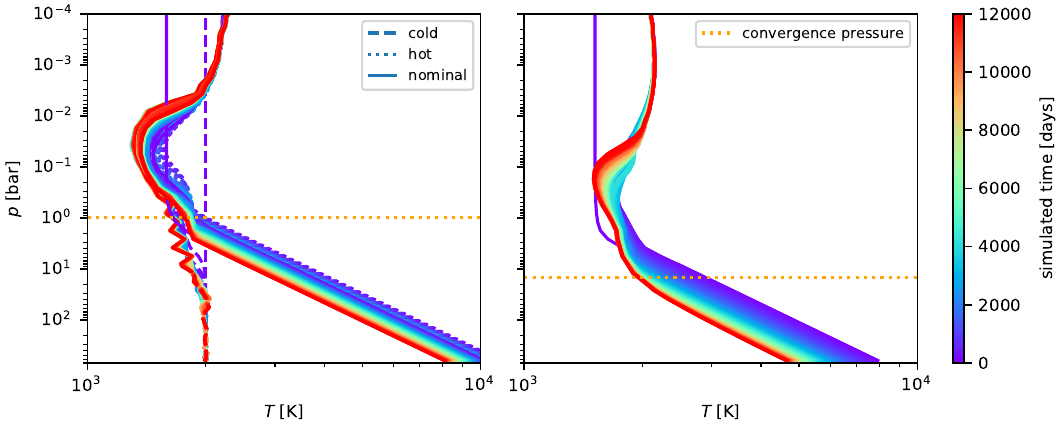}
        \caption{Evolution of the temperature profile at the substellar point for HD~209458b (left) and WASP-43b (right). The left panel includes the evolution of two additional HD~209458b simulations with a colder and a hotter initial temperature profile. The right panel includes one additional WASP-43b simulation with a cooler initial temperature profile. The additional simulations initialized with an adiabat ran for 3000 days, whereas the cool HD~209458b ran for 5000 days. The temperature time averaging and output interval is set to 100 days, therefore not resolving the fast semi-convergence of the temperature in the photosphere during the first 100 days. The horizontal orange dashed line illustrates the approximate altitude at which radiative heating or cooling becomes very small.}
        \label{fig:temp_evol}
\end{figure*}
The GCM experiments of \citet{Sainsbury-Martinez2019Inflation} indicate that hot Jupiters might pump the energy input from irradiation into the deep atmosphere ($p>\SI{1}{\bar}$) by means of dynamic heat transport, leading to hot temperatures in their deep atmosphere. The equation of entropy conservation in steady state can be written as \citep[e.g.,][]{Tremblin2017Inflation, Sainsbury-Martinez2019Inflation}
\begin{equation}\label{eq:entropycons}
        \frac{\mathrm{d}\Theta}{\mathrm{d}t} - \boldsymbol{u}\cdot\nabla\Theta = 0,
\end{equation}
where $\boldsymbol{u}$ is the velocity vector, $\nabla\Theta$ is the spatial gradient of the potential temperature, $\Theta,$ and $\frac{\mathrm{d}\Theta}{\mathrm{d}t}$ encapsulates dissipative process that cool or heat the atmosphere\footnote{$\frac{\mathrm{d}\Theta}{\mathrm{d}t}$ is given by Eq.~\ref{eq:energyeq} in this work, but other processes such as Ohmic dissipation could also play a role.}. If we only consider irradiation, $\frac{\mathrm{d}\Theta}{\mathrm{d}t}$ becomes very small in the deep layers of the atmosphere because most of the stellar light is absorbed in the upper atmosphere. However, this requires that either $\boldsymbol{u}$ or $\nabla \Theta$ become very small. To fulfill the latter, the potential temperature must be close to constant and therefore follow an adiabat ($T \propto p^\eta$).

The potential temperature of such an adiabat would therefore be mainly determined by the temperature of the convergence pressure, which is defined as the region where heating and cooling by radiation becomes very small. Furthermore, the convergence pressure lies higher up in the atmosphere compared to the pressure at which the atmosphere would be convective.

The weakness of radiative cooling and heating in the deep layers of the atmosphere requires long model runtimes. We can utilize the fast runtime of \texttt{expeRT/MITgcm} of 600-1000 simulated days per day of model runtime to study the temperature convergence in the deeper layers of the atmosphere. We investigate the temperature evolution of HD~209458b and WASP-43b in Fig.~\ref{fig:temp_evol}, where we show the temperature profile at the substellar point as a function of time. For HD~209458b, we show two additional simulations with different initial temperature profiles. The first one is identical to the nominal model, but uses $\Theta_\mathrm{ad}=\SI{2000}{K}$ for the initial deep adiabat, instead of $\Theta_\mathrm{ad}=\SI{1800}{K}$ for the nominal model (see Sect.~\ref{sec:ic}); whereas the second simulation was initialized with a globally isothermal temperature of $T=\SI{2000}{\K}$. We performed a similar test for WASP-43b, where we ran a model of WASP-43b that has been initialized, with $\Theta_\mathrm{ad}=\SI{900}{\K,}$ instead of the nominal $\Theta_\mathrm{ad}=\SI{1400}{\K}$. We chose to stop the extra HD~209458b and WASP-43b simulations with an adiabatic interior at $t=\SI{3000}{\day}$, since the only purpose of these simulations is to gain information about the dependence of the temperature convergence behavior on initial conditions. However, it is important to note that the convergence time needed to cool down the deep atmosphere from a hot state to a colder state is shorter than the time needed to heat the deep atmosphere from a cold state to a hotter state \citep{Sainsbury-Martinez2019Inflation}.

We find that for WASP-43b, the temperature in the deep layers is steadily evolving during the full \SI{12000}{\day} simulation. The rate of temperature change drops from approximately $\SI{1.5}{\K\per\day}$ to approximately $\SI{0.05}{\K\per\day}$ at the end of the simulation, while slowly converging. This process of cooling results in an almost isothermal temperature profile in the photosphere of WASP-43b, which is subsequently changing into a more pronounced temperature inversion once the deep cooling becomes less efficient (see Fig.~\ref{fig:temp_evol}, right panel). The decaying rate of cooling in the deep layers forecasts a final cool adiabat of $\Theta_\mathrm{ad}\approx\SI{700}{\K}$.

The temperature evolution of the initially cool WASP-43b simulation overlaps with the temperature evolution of the nominal simulation. However, the later states of the nominal WASP-43b including the thermal inversion matches the initially cold simulation fairly well. We therefore find that the evolution of the temperature of the initially cold WASP-43b simulation confirms the independence of the final state of the deeper atmosphere of WASP-43b on initial conditions.

HD~209458b, on the other hand, appears to maintain its initial temperature profile in the deep layers (see Fig.~\ref{fig:temp_evol}, left panel). The temperature constantly decreases with a rate varying around $\SI{0.1\pm0.1}{\K\per\day}$. Furthermore, we see that the deep temperature evolution is similar for the hot initial temperature profile, yielding very low values for the temperature decay in time. A perfect initial guess by chance is therefore impossible. We therefore conclude that we could not reach temperature convergence in the deep layers of HD~209458b within 12000 days. This outcome, while \enquote{negative,} is of significance for other studies that have already published temperature maps of HD~209458b.

The temperature in the isothermally initialized HD~209458b simulation does not evolve towards a hotter state in the deep layers. It is not possible to predict whether the final state of the temperature in the deep layers of these simulations will indeed be adiabatic. We can thus not confirm \citet{Sainsbury-Martinez2019Inflation} in that the deep layers of this simulation heat up towards such a state. Additionally, the temperature exhibits a shaky pattern that is likely linked to the deeper parts of the atmosphere. This pattern arises at the convergence pressure and could be explained by upward propagating gravity waves; a process that has also been postulated for the atmospheres of brown dwarfs \citep[e.g.,][]{Freytag2010Convection, Showman2013Dwarfs}. We therefore postulate that this pattern is linked to the strength of the dynamic processes in the deeper parts of the atmosphere. It remains an open question if this pattern disappears when convergence is reached or if it could be forced to a stably stratified temperature profile with the use of a convective adjustment scheme \citep[e.g.,][]{Deitrick2020GCM,Lee2021GCM}, which we do not use within this work. We plan to examine these gravity waves and the effect of a convective adjustment scheme in detail in an upcoming work.

In the following, we investigate the effect of surface gravity and planetary rotation on the convergence behavior of the temperature in the deep atmosphere. Radiative timescales scale reciprocally with the surface gravity \citep[see e.g.,][]{Showman2020review}, leading to longer radiative timescales in HD~209458b at similar pressure levels compared to WASP-43b, even though both planets receive a similar amount of energy from their host star. The pressure level where the radiative heating and cooling becomes irrelevant is therefore different for both planets, where irradiation is penetrating into deeper pressure layers in the atmosphere of WASP-43b compared to HD~209458b. This will then influence the pressure level and temperature at which a deep adiabat could couple to the radiatively dominated part of the atmosphere \citep[according to the explanations of][]{Tremblin2017Inflation,Sainsbury-Martinez2019Inflation}. Hence, we think that a colder interior of WASP-43b in our simulations is a natural consequence of the assumption of the value of the surface gravity.

\begin{figure}
        \centering
        \includegraphics[width=.45\textwidth]{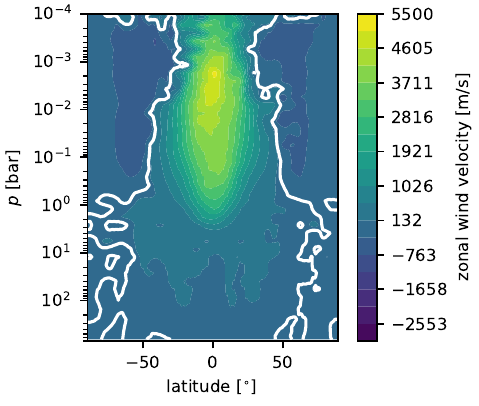}
        \caption{Zonally averaged wind velocities (like Fig.~\ref{fig:WASP-43b_zonal_mean}) for a model of WASP-43b with a rotation period equal to the rotation period of HD~209458b.}
        \label{fig:WASP-43b_long_zonal_mean}
\end{figure}
We performed an additional simulation of a WASP-43b-like planet with a rotation period equal to the rotation period of HD~209458b ($P_\mathrm{rot}=\SI{3.47}{\day}$), instead of the nominal short rotation period of $P_\mathrm{rot}=\SI{0.8135}{\day}$. The goal of this additional simulation is to consider the dependence of the temperature evolution on rotation. The zonal wind velocity in the case of slow rotation is significantly different to our nominal model of WASP-43b, where we find that unlike the nominal simulation (Fig.~\ref{fig:WASP-43b_zonal_mean}), the slowly rotating WASP-43b simulation does not exhibit a deep equatorial jet (Fig.~\ref{fig:WASP-43b_long_zonal_mean}). This finding agrees well with the work of \citet{Carone2020GCM}, who predicted that WASP-43b exhibits a deep superrotating jet, which is linked to the fast rotation of WASP-43b.

\begin{figure}
        \centering
        \includegraphics[width=.45\textwidth]{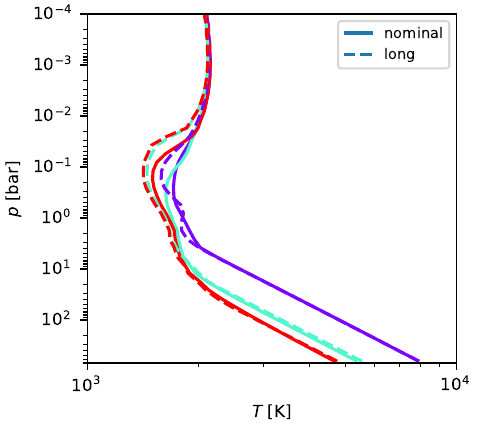}
        \caption{Evolution of the temperature profile at the substellar point (like Fig.~\ref{fig:temp_evol}). Three time steps (\SIlist{100;5000;12000}{\day}) of the nominal model of WASP-43b are compared to a model of WASP-43b with a rotation period equal to the rotation period of HD~209458b. The color scale is the same as in Fig.~\ref{fig:temp_evol}.}
        \label{fig:long_vs_short}
\end{figure}

Recently, \citet{Sainsbury-Martinez2021Inflation} proposed that the efficiency of downward heat transport of energy depends on the presence or absence of rotational dynamics over advective dynamics. If rotational dynamics dominate over advective dynamics, which is the case for a deep equatorial jet, we would expect a lower efficiency for the downward heat transport of energy. We therefore compared the temperature in the model of WASP-43b with a deep jet (nominal) to the model without a deep jet ($P_\mathrm{rot}=\SI{3.5}{\day}$). We show the resulting substellar point temperature profiles in Fig.~\ref{fig:long_vs_short}, where we find that both simulations exhibit a similar temperature evolution in the deep atmosphere. The independence of the deep temperature profile on the rotation period leads us to the conclusion that the fast temperature convergence of WASP-43b can be almost exclusively explained by the continuously strong radiative heating and cooling in the deep atmosphere caused by the high value of surface gravity.

The requirement of zero radiative heating and cooling (see Eq.~\ref{eq:entropycons}) for the theory of downward transport of energy \citep{Sainsbury-Martinez2019Inflation} can only be monitored with a climate model that includes radiative feedback. Using our model, we find that radiative heating and cooling can still be important at greater depths, bringing on the question of how the requirements for the formation of a hot deep adiabat might be matched. However, we do note that our simulations demonstrate that our non-gray GCM succeeds in maintaining an adiabatic deep atmosphere, without the need for a convective adjustment scheme. We propose that a further detailed analysis is needed to settle the question of whether a downward transport of energy is possible.

\section{Synthetic spectra}\label{sec:spectra}
\begin{figure}
        \begin{subfigure}{.45\textwidth}
                        \includegraphics[width=\textwidth]{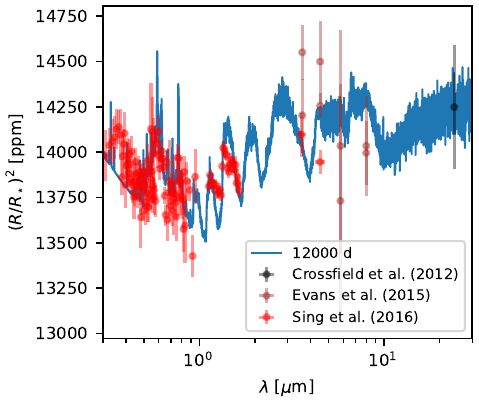}
        \end{subfigure}
        \begin{subfigure}{.45\textwidth}
                \includegraphics[width=\textwidth]{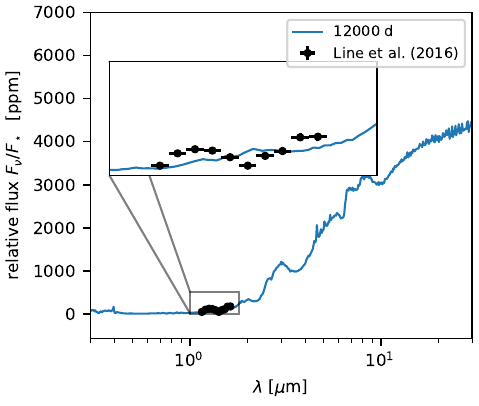}
        \end{subfigure}
        \caption{Synthetic transmission spectrum (top) and dayside emission spectrum (bottom) for HD~209458b calculated from simulation results after 12000 days, compared to the observations. The thermal inversion, caused by the TiO and VO opacities, overcasts the observed water feature at $\SI{\approx 1.4}{\mu\m}$.}
        \label{fig:spec_HD2}
\end{figure}
\begin{figure}
        \begin{subfigure}{.45\textwidth}
                        \includegraphics[width=\textwidth]{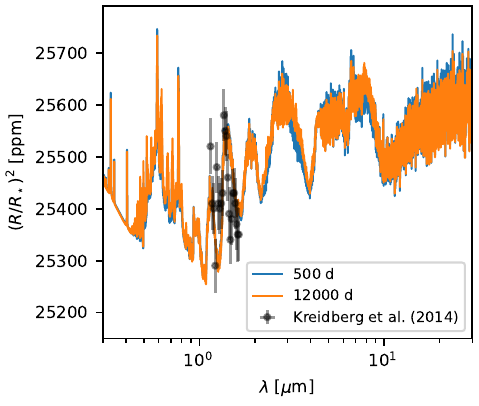}
        \end{subfigure}
        \begin{subfigure}{.45\textwidth}
                \includegraphics[width=\textwidth]{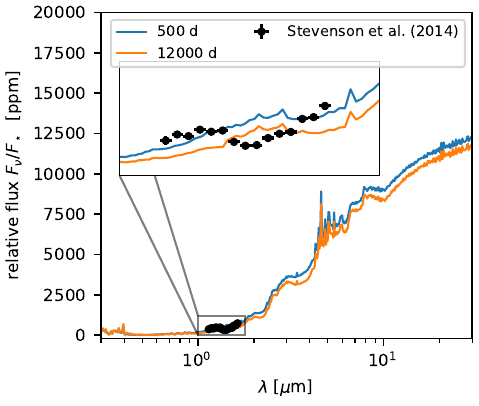}
        \end{subfigure}
        \caption{Synthetic transmission spectra (top) and dayside emission spectra (bottom) for WASP-43b calculated from simulation results after 500 and 12000 days, compared to the observations. The thermal inversion, caused by the TiO and VO opacities, overcasts the observed water feature at $\SI{\approx 1.4}{\mu\m}$.}
        \label{fig:spec_WASP-43b}
\end{figure}

We post-processed our GCM results using \texttt{petitRADTRANS} to obtain emission and transmission spectra. We followed \citet{Baeyens2021GCM} and calculated the transmission spectrum using the average temperature in latitude $\pm 20^\circ$ around the west and east terminators ($\varphi=-90^\circ$ and $\varphi=90^\circ$). Such an isothermal treatment of the terminator region is possible since temperature variations in the terminator region are very small for pressure layers below \SI{1e-2}{\bar}, which are most relevant to the transmission spectrum. The transmission spectrum is then calculated using the radius and gravity values of Table~\ref{tab:param} for a reference pressure of \SI{0.01}{\bar}. As in \citet{Baeyens2021GCM}, we determined the final transit spectrum by quadratically averaging the east and west terminators. For WASP-43b, we show the \citet{Kreidberg2014Clouds} HST/WFC3 data around the $\SI{\approx 1.4}{\mu\m}$ H$_2$O feature, whereas for HD~209458b, we show the data of \citet{Crossfield2012HD20,Sing2016HD20,Evans2015HD20}. We calculate the mean of the observed values of the wavelength dependent radius as well as of the calculated transmission spectra for each observed spectral window to determine the offset in radius between observations and our model. We then corrected the observed radius by subtracting this offset. The offset correction is necessary due to differences in the normalization of the radius \citep[see e.g.,][]{Lee2021GCM}.

For the emission spectrum, we chose to regrid our GCM output onto a low resolution rectangular grid with a 15$^\circ$ longitude-latitude resolution (288 horizontal grid cells). We then calculated the angle dependent planetary intensity at the top of each vertical column using a derivative of the flux routine of \texttt{petitRADTRANS} for 20 Gaussian $\mu$ quadrature points. We then integrated the intensity field over the solid angle to obtain the flux in the direction of the observer \citep[see e.g.,][]{Seager2010rev, Amundsen2016UKMetGCM}. The general procedure is identical to the one used in \citet{Carone2020GCM} and only differs in its numerical implementation, where we use \texttt{petitRADTRANS} instead of \texttt{petitCODE} to obtain the intensity.\footnote{The python package used to obtain and integrate the emission spectrum has been made publicly available and can be found at \url{https://prt-phasecurve.readthedocs.io/}.} We tested the impact of the low-resolution spatial grid on the resulting emission spectrum by testing an even lower spatial resolution of 30$^\circ$ (72 grid cells) and found that only minimal differences occur, validating the 288 grid cells approach outlined above. We chose to plot the data of \citet{Stevenson2014WASP} for WASP-43b and the data of \citet{Line2016HD20} for HD~209458b. We note that we do not correct the observed emission spectrum for systematic errors, but instead plot the published values.

We investigate whether the difference in the magnitude of the temperature inversion caused by the ongoing convergence of the deeper atmosphere of our WASP-43b simulation would impact predictions for observables. The resulting transmission spectra for HD~209458b in Fig.~\ref{fig:spec_HD2} and WASP-43b in Fig.~\ref{fig:spec_WASP-43b} show good agreement with the observed values. However, we find that the pressure region between around \SI{1}{\bar} contributes most to the spectrum for wavelengths between $\SI{0.4}{\mu\m}$ and $\SI{2}{\mu\m}$, whereas the pressure range between \SI{0.1}{\bar} and \SI{0.001}{\bar} dominates the infrared spectrum. This would explain, why we only see a (very small) difference between the transmission spectra for WASP-43b at \SI{500}{\day} and \SI{12000}{\day} in the infrared but not in the optical (see Fig.~\ref{fig:spec_HD2}).

In examining the dayside emission spectra, we find that the shape of the water line feature of WASP-43b and HD~209458b does not seem to agree between the observed water feature and the model spectra. The reason for the difference in the shape of the water feature is found in the inclusion of TiO and VO opacities, which cause a temperature inversion \citep{Showman2009GCM}. A similar trend can be found in \citet[][Fig. 7, right panel]{Lee2021GCM}, where the authors show the spectral differences of a semi-gray model without temperature inversion compared to more advanced models that include the temperature inversion. A model without TiO and VO in the upper atmosphere would therefore clearly fit the observed spectra much better. Further, the lack of clouds and the assumption of solar metallicity in our model affects the emitted flux and also effect the strength of superrotation \citep[e.g.,][]{Kataria2015WASP, Drummond2018Metallicity, Parmentier2021GCM}.

Moreover, we find that the \SI{500}{\day} and \SI{12000}{\day} emission spectra of the WASP-43b model are clearly impacted by the state of the deep atmosphere (see Fig.~\ref{fig:temp_evol}). The flux differences in the emission spectrum thus highlight the effect of deep temperature convergence on the observable atmosphere. These differences are not seen for HD~209458b because the observable atmosphere seems to be unaffected by the state of the deep atmosphere (see Fig.~\ref{fig:temp_evol}).

\section{Discussion}\label{sec:discussion}

In this work, we employ \texttt{expeRT/MITgcm}, a new 3D GCM with non-gray radiative transfer specifically built to investigate the processes at work in greater depths in more detail. The first application of this model to WASP-43b and HD209458b shows interesting results in flow and temperature at such depths and may thus shed light onto the unsolved question of hot Jupiter inflation.

\subsection{Versatile fast 3D GCM with non-gray radiative transfer}

There are currently several 3D exoplanet atmosphere models available, which vary in their complexity of radiative transfer \citep{Showman2009GCM,Rauscher2012double,DobbsDixon2013GCM,Amundsen2016UKMetGCM,Mendonca2018GCM,Lee2021GCM} and even in the basic implementation of the Navier Stokes equation to treat fluid dynamics, ranging from hydrostatic primitive equations \citep[e.g.,][]{Showman2009GCM} to full \citep[e.g.,][]{Mayne2014UM,Mendonca2018GCM,Deitrick2020GCM}. In addition, several 3D climate models include additional physical processes, which are thought to be important for the understanding of exoplanet atmospheres: clouds \citep{Lee2016Clouds,Lines2019Clouds,Parmentier2021GCM,Roman2021Clouds}, photochemical hazes \citep{Steinrueck2021haze}, disequilibrium chemistry \citep{Agundez2014GCMChem,Drummond2018GCMChem,Mendonca2018Obs,Baeyens2021GCM}, and magnetic fields \citep{Rogers2014MHD}. Cloud modeling in 3D climate models ranges from detailed kinetic modeling \citep[e.g.,][]{Lee2016Clouds} to simple but efficient parametrization \citep[e.g.,][]{Roman2021Clouds,Christie2021Clouds}.

In addition to cloud modeling, disequilibrium chemistry, and magnetic field interaction, there is a need for GCMs that are capable of investigating processes at greater depths such as deep wind flow and deep energy transport. In this work, we introduce a new non-gray GCM formalism that is numerically efficient and thus ideally suited to tackle the challenging processes at greater depths. The main reason for these advances is the use of the Feautrier method in combination with accelerated lambda iteration to solve the radiative transfer equation (Sects.~\ref{sec:RT_flux} and \ref{sec:source_convergence}). We find that these methods allow us to run non-gray GCM models for 1000 simulation days within just a single day of run time and using 24 cores. Applying the code to WASP-43b and HD~209458b shows that the upper atmosphere agrees well with the results of other non-gray GCMs (Sect.~\ref{sec:results}) for these planets \citep{Kataria2015WASP,Showman2009GCM,Amundsen2016UKMetGCM,Lee2021GCM}. The differences with Newtonian cooling as used in \citet{Carone2020GCM} are shown in Appendix.~\ref{sec:newton}. The computational efficiency of \texttt{expeRT/MITgcm} allows extending the work of \citet{Sainsbury-Martinez2019Inflation} to investigate temperature convergence for very long simulation times at deeper layers without artificial external forcing mechanisms (Sect.~\ref{sec:temp_convergence}).

\subsection{Importance of the temperatures in the deep atmosphere}
The temperature in layers below \SI{10}{\bar} can influence the observable chemistry in several ways: for instance, iron and magnesium could condense in these layers and consequently would not be present in the gas phase of the upper atmosphere, where they would be observable via high-resolution spectroscopy \citep[][Fig.~13]{Sing2019observ}. Other possibilities include the quenching of methane via disequilibrium chemistry \citep{Agundez2014DeepQuench,Baeyens2021GCM} and the thickness of clouds also depends on the temperature at deeper layers \citep{Fortney2020Interior}. Silicate cloud formation can remove or inject oxygen-bearing molecules from the gas phase \citep{Helling2019bClouds,Helling2021CloudsParam}. Therefore, deep cloud condensation might affect local C/O ratios in cloud forming regions even more than thinner clouds. At the same time, the condensation of important cloud particles such as magnesium at greater depths may affect the composition of clouds higher up, as pointed out by \citet[]{Helling2021CloudsParam}.

In this work, we focused on energy transport and thermal evolution at greater depths, because it is still an open question whether energy can be injected via 3D circulation from the irradiated photosphere into deeper layers in hot to ultra-hot Jupiters that could at least partly explain hot Jupiter inflation \citep{Mayne2014UM, Tremblin2017Inflation,Mayne2017GCM}. For WASP-43b, we find that higher surface gravities compared to HD~209458b already  naturally lead to a deeper convergence pressure. Thus, setting the surface gravity and assuming long-enough run times drives the simulation at greater depths naturally to a colder state. The surface gravity of WASP-43b, however, is higher because it is not inflated. In other words: the state of inflation sets the surface gravity, which, in turn, strongly affects the convergence behavior towards the final state. This dependence of the convergence behavior on the state of inflation is obviously not optimal, since it creates a \enquote{chicken-and-the-egg problem} when investigating energy transport at greater depths. However, we stress that our numerical efficient non-gray GCM has the potential to explore this problem in more detail, where we will also take additional dissipative processes at greater depths into account in the future.

\subsection{Effect of deep dynamics on the observable atmosphere}

Only a few GCM studies of hot Jupiters have been conducted thus far with the aim to explore dynamical effects at greater depths owing to the computational challenges that need to be resolved in the deeper atmosphere. \citet{Mayne2017GCM} explored the possibility of deep wind flow when horizontal temperature differences at greater depths are imposed. \citet{Carone2020GCM} showed that fast rotators like WASP-43b (orbital periods less than \SI{1.5}{\day}) naturally evolve equatorial jets that can extend very deep into the interior. In the Solar System, the fast-rotating gas giants are also shown to have deep jets \citep{Kaspi2018DeepJetJupiter,Iess2019}. In this work, the deep equatorial jets in WASP-43b and shallow equatorial jets in HD~209458b are reproduced. Thus, we stress, as does \citet{Carone2020GCM}, that some extrasolar planets like the JWST ERS target WASP-43b \citep{Venot2020chem,Bean2018JWST} can be dynamically active at greater depth greater than \SI{200}{\bar}, which is the typical lower boundary assumed for hot Jupiter GCMs \citep[e.g.,][]{Showman2009GCM,Amundsen2016UKMetGCM,Deitrick2020GCM,Parmentier2021GCM,Lee2021GCM}. More importantly, we find in this work that resolving dynamics at greater depths is important to allow the deep atmosphere of WASP-43b to self-consistently evolve towards cool adiabat ($\Theta_\mathrm{ad}\approx 700$~K).

It is not only the energy transport at greater depths that is influenced by deep dynamics. Previous works indicate that the presence of a deep flow may impact the observable wind and temperature structure of hot Jupiters. \cite{Mayne2017GCM} showed that the deep wind flow can greatly diminish superrotation in the observable atmosphere. \citet{Wang2020GCM} found that the simulated wind flow structure in their mini-Neptune simulation can abruptly change from two jets to one jet, which could be observable via changes in the hotspot location. This change happens without forced conversion after 50000 simulation days. Furthermore, \citet{Carone2020GCM}  find that in their simulations, superrotation can be completely disrupted at the dayside, leading to a local retrograde flow. In this work, we do not find this retrograde flow, but instead unperturbed superrotation. This can partly be attributed to the fact that our models do not rely on artificial temperature forcing in the deep atmosphere, but instead, they can capture the feedback between dynamics and irradiation that tends to strengthen superrotation. \citet{Carone2020GCM} also noted that superrotation is never completely removed, and that there are apparently two tendencies at play in the upper atmosphere: one leading to superrotation and one leading to other flows.

\citet{Parmentier2021GCM} noted that the strength of superrotation in the observable atmosphere at the dayside can be diminished by nightside cloud coverage. \citet{Beltz2022Interior, Hindle2021MHD} indicated that magnetic field interaction can completely suppress superrotation. Observationally, it was found that CoRoT-2b appears to have a retrograde flow at the dayside \citep{Dang2018CoRoT2b} and HAT-P-7b may even have a dynamically shifting hot spot \citep{Armstrong2016Hat-P-7b}. Thus, future works that includes large-scale flow at greater depths is warranted in order to investigate the interplay among more complex physics and its effect on the observable atmosphere.

It has been further hypothesized that shocks, mechanical dissipation of strong winds, and turbulent vertical mixing at greater depths may also inject energy into the interior and contribute to inflation \citep{Li2010MechanicDissipation, Heng2012Shock, Perna2012Shocks, Fromang2016Shock, Menou2019VerticalMixing}. It is, however, unclear whether shear flow instabilities at greater depths \citep{Menou2009Shear,Rauscher2010GCM,Liu2013LowerBoundary,Carone2020GCM} may affect the vertical heat flow. Tackling these processes requires higher resolution non-hydrostatic 3D atmosphere simulations \citep{Menou2019VerticalMixing,Fortney2021review}.

We note that the question of the importance of dynamical processes at greater depths in atmospheres is neither new nor solely constrained to hot Jupiters. There are clear parallels between the dynamics and the heat transport at greater depths in brown dwarfs \citep{Sainsbury-Martinez2021Inflation}. Considerable work has also been done in understanding such processes both for the Earth's climate and for stellar atmospheres. Such examples are clumpy dust formation triggered by shock waves via large scale convective flows at greater depths in AGB stars \citep{Fleischer1992AGBcloudClumps, Woitke2005AGBcloudClumps,Hoefner2019AGBcloudClumps} and the need to resolve downdrafts reaching deep into the interior \citep{Freytag2017ConvectionAGB,Freytag20193DAGB} or \enquote{convective overshooting} at the boundary of convective and stably stratified regions in stars \citep[e.g.,][]{Hanasoge2015SunReview,Bressan1981MassiveOvershoot}. The issue of \enquote{convective overshooting} seems to resemble the boundary problem between a convective interior and a stably stratified irradiated atmosphere in hot Jupiters. Similar processes are also relevant in Earth GCMs: the importance of dry and moist convection in the Earth's atmosphere and surface friction was identified early on \citep{Manabe1965EarthConvection,Peixoto1984EarthClimate}. The importance of convection and surface friction was also identified in GCMs of tidally locked exo-Earths \citep{Carone2016SurfaceExoEarth,Koll2016ExoEarthSurface,Sergeev2020ConvectionExoEarth} and for brown dwarfs \citep{Tremblin2019LowerBoundary}. Thus, it comes as no surprise that processes at greater depths matter in hot Jupiters as well. More importantly, deep dynamical processes impact the large-scale circulation higher up and can potentially yield important insights into the evolution of exoplanets.

\section{Summary and conclusions}
\label{sec:conclusion}

Here, we introduce and explain \texttt{expeRT/MITgcm}, which extends the deep wind framework \texttt{exorad/MITgcm} of \citet{Carone2020GCM} with full gaseous radiative transfer, including scattering. We demonstrate that the Feautrier method with accelerated lambda iteration and correlated-k is a fast and accurate approach to model the non-gray radiative transfer while still achieving model runtimes of hundreds of simulation days in a single runtime day.

We simulated the two hot Jupiters HD~209458b and WASP-43b which are almost equally irradiated, but which have different orbital periods ($P_\mathrm{rot}^\mathrm{HD~209458b}=\SI{3.47}{\day}$ and $P_\mathrm{rot}^\mathrm{WASP-43b}=\SI{0.8135}{\day}$) and surface gravities ($g^\mathrm{HD~209458b}=\SI{8.98}{\m\per\s\squared}$ and $g^\mathrm{WASP-43b}=\SI{46.9}{\m\per\s\squared}$).

We find that our results for the observable atmosphere of HD~209458b and WASP-43b agree well with the findings of previous works. Our WASP-43b models were able to confirm a deep equatorial jet as well as a dependency of the temperature in the observable atmosphere on the state of the deep atmosphere as predicted by \citet{Carone2020GCM}. However, we did not find the retrograde equatorial flow. Future work is needed to investigate the reason for these differences.

We investigated the deep atmosphere (pressures between \num{1} and \SI{700}{\bar}) of WASP-43b and HD~209458b and found a difference in the temperature evolution. WASP-43b cools down to a cold temperature in the deep atmosphere, whereas HD~209458b successfully maintains its initial temperature profile within the runtime. This difference in the temperature evolution can be linked to the different values of the surface gravity, where we find that due to the high surface gravity of our WASP-43b simulations, radiative heating and cooling are very important at greater depths. We looked at the dependency of the temperature evolution on the rotation period and found that the difference in temperature evolution is not significantly enhanced by the presence or absence of a deep equatorial jet.

Last but not least, we caution that longer convergence time scales of at least 10000 simulation days are needed to properly resolve the observable atmosphere temperature structure of non-inflated exoplanets such as WASP-43b. The observable atmosphere of inflated hot Jupiters such as HD~209458b seems to be decoupled from the deep atmosphere and, therefore, it can be expected to be resolved after 1000 days of simulation time.

More work is warranted to identify whether and how energy transport from the irradiated atmosphere into the interior contributes to inflation. The computationally efficient non-gray GCM \texttt{expeRT/MITgcm} that we introduce here is an ideal tool to self-consistently tackle deep wind flow and energy transport.

\begin{acknowledgements}
A.D.S., L.D., U.G.J, S.K. and C.H. acknowledge funding from the European Union H2020-MSCA-ITN-2019 under Grant no. 860470 (CHAMELEON). L.C. acknowledges support from the DFG Priority Programme SP1833 Grant CA 1795/3. L.D. acknowledges support from the KU LEUVEN interdisciplinary IDN grant IDN/19/028 and the FWO research grant G086217N. U.G.J acknowledges funding from the Novo Nordisk Foundation Interdisciplinary Synergy Program grant no. NNF19OC0057374. P.M. acknowledges support from the European Research Council under the European Union’s Horizon 2020 research and innovation program under grant agreement No. 832428-Origins. R.B. acknowledges support from a PhD fellowship of the Research Foundation -- Flanders (FWO).
The computational resources and services used in this work were provided by the VSC (Flemish Supercomputer Center), funded by the Research Foundation Flanders (FWO) and the Flemish Government – department EWI.
We also thank an anonymous referee for their very useful comments that helped a lot in the interpretation of the obtained results.
\end{acknowledgements}

\bibliographystyle{aa}
\bibliography{ms}

\begin{appendix}

\section{Importance of sphericity in stellar attenuation}\label{sec:planeparallel}
The incident angle for the stellar light onto the planetary atmosphere is accounted by the factor $\mu_\star$. In Eq.~\ref{eq:mu_star_definition}, $\mu_\star$ is derived assuming plane-parallel geometry, which is a broadly used approximation \citep[see e.g.,][]{Showman2009GCM, Amundsen2016UKMetGCM, Lee2021GCM}. Because this paper is meant to benchmark the radiative transfer implementation by comparing it with previous work, we used plane-parallel atmospheres and did not account for the sphericity of the model atmospheres.

While the incident angle in the substellar region is independent of depth and agrees well with the plane-parallel assumption, substantial differences are expected at the polar regions, where the effective solar path length deviates significantly from the solar path length in plane-parallel, ultimately altering the angle of incidence. The expected deviation in $\mu_\star$ between a spherical and a plane-parallel approximation increases with pressure (and, therefore, optical depth) and is potentially more important at greater depths. On the other hand, we note that the stellar component of the flux $F^\star$ decreases with optical depth.

A correct treatment of sphericity would involve rewriting Eqs.~\ref{eq:J_ini} and \ref{eq:tau}, which would lead to a double integral over the solid angle. Since this is not computationally feasible, some authors \citep{Li2006PP, Mendonca2018Obs} who account for sphericity in the environment of a GCM have expanded the plane-parallel assumption with the correction of the effective path-length. The resulting incident angle $\mu_{\star,\mathrm{eff}}(z)$ would then be dependent on the height $z$ and the planetary radius $R_p$ and is in its simplest approximation given by \citep{Li2006PP}:
\begin{equation}
        \mu_{\star,\mathrm{eff}}(z)=\sqrt{1-\left(\frac{R_p}{R_p+z}\right)^2(1-\mu_{\star}^2)}.
\end{equation}

In conclusion, the plane-parallel assumption is suitable if only the uppermost part of the atmosphere is probed or if the error of the optical depth within the polar regions causes only negligible effects. If the atmosphere is to be probed deeper, an additional treatment of the sphericity is vital \citep[e.g.,][]{Li2006PP}. We plan to incorporate this aspect in an upcoming work.

\section{Verification of the radiative transfer}\label{sec:tests}
It has been shown in previous works \citep{Showman2009GCM,Amundsen2016UKMetGCM,Lee2021GCM} as well as in this work that the inclusion of a full gaseous radiative transfer in GCMs is vital to obtain a full understanding of the atmospheric dynamics of hot Jupiters. However, a correct treatment of radiative transfer in the framework of a GCM is computationally very expensive and simplifications need to be well understood and justified. The correlated-k approximation and the assumption of equilibrium chemistry both lead to significant simplifications.

We tested the validity of radiative transfer in our model in a three-step process. We first show a comparison of the fluxes and heating rates obtained with \texttt{petitRADTRANS} and \texttt{expeRT/MITgcm} in Sect.~\ref{sec:tests_am} and then our testing of the wavelength binning, done by comparing the resulting temperature profile in Sect.~\ref{sec:S0vsS1}. We finally show our testing of the influence of the horizontal interpolation of fluxes on the temperature profile in Sect.~\ref{sec:interp_vs_nointerp}.

\subsection{Verification of fluxes and heating rates}\label{sec:tests_am}
\begin{figure*}
        \centering
        \includegraphics[width = \textwidth]{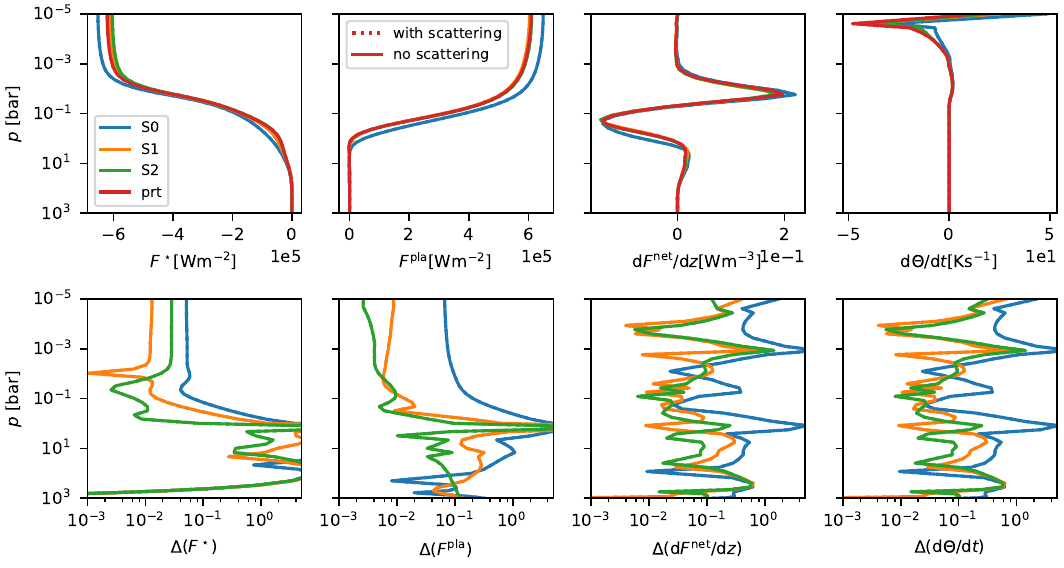}
        \caption{Test 3 of \citet{Amundsen2014GCM} for a mixed dayside hot Jupiter atmosphere, where we compare fluxes and heating rates for different wavelength resolutions (see Table~\ref{tab:opac_bins_overview}) in \texttt{expeRT/MITgcm} (S0, S1, S2) to \texttt{petitRADTRANS} (prt). The bottom panels display the residuals of the above panels. From left to right, we show the incoming stellar flux, the emitting planetary flux, the gradient of the net flux, and the resulting change of potential temperature.}
        \label{fig:test3}
\end{figure*}

We follow the suggestion of \citet{Amundsen2014GCM} and implement their dayside test for a cloud-free atmosphere in \texttt{expeRT/MITgcm}. We show the fluxes and heating rates in Fig.~\ref{fig:test3}, where we compare the results obtained with \texttt{expeRT/MITgcm} in different wavelength resolutions with scattering turned on and off to results obtained with \texttt{petitRADTRANS}, which has been benchmarked in \citet{Baudino2017}.

We find that our results agree well with \texttt{petitRADTRANS}, which uses a much higher wavelength resolution of approximately 1000 times more correlated-k wavelength bins compared to the nominal resolution of our models (see Table~\ref{tab:opac_bins_overview}). We note that the employed method of random sampling to combine opacities of individual species is prone to errors in the order of a few percent. On top of that, there is a general loss of accuracy when using low-resolution opacities \citep[e.g.,][]{Leconte2021exok}, which is also in the order of a few percent.

The differences in the bolometric stellar flux ($F^\star_\nu$) can be explained by the numerical accuracy of the integration, which depends on the amount of frequency bins used to integrate the spectral flux (see Eq.~\ref{eq:F_star}). Reducing this error would require calculations of the stellar bolometric flux for a high wavelength resolution and a normalization of the stellar spectral flux to the high-resolution bolometric flux. The current low-resolution implementation of \texttt{petitRADTRANS} does not include such a normalization, so any future implementation of \texttt{expeRT/MITgcm} might improve on that.

Overall, there seems to be a general  good agreement between Fig.~\ref{fig:test3} and the simulations shown in \citet{Amundsen2014GCM}. We note that \citet{Amundsen2014GCM} used the Kurucz stellar spectrum\footnote{see \url{http://kurucz.harvard.edu/stars/hd209458/}}, while we used the PHOENIX stellar spectrum \citep{Husser2013PHOENIX} that is part of \texttt{petitRADTRANS}. Differences might also occur due to the different opacities, where we used the same species as mentioned in \citet{Amundsen2014GCM}, but from updated opacity sources (see Table~\ref{tab:opac}). Furthermore, we used the equilibrium chemistry of \citet{Molliere20191Dmodel}, while \citet{Amundsen2014GCM} used the analytic abundances of \citet{BurrowsSharp1999Opac}.

\subsection{S0 vs S1}\label{sec:S0vsS1}
\begin{figure*}
        \centering
        \includegraphics[width=\textwidth]{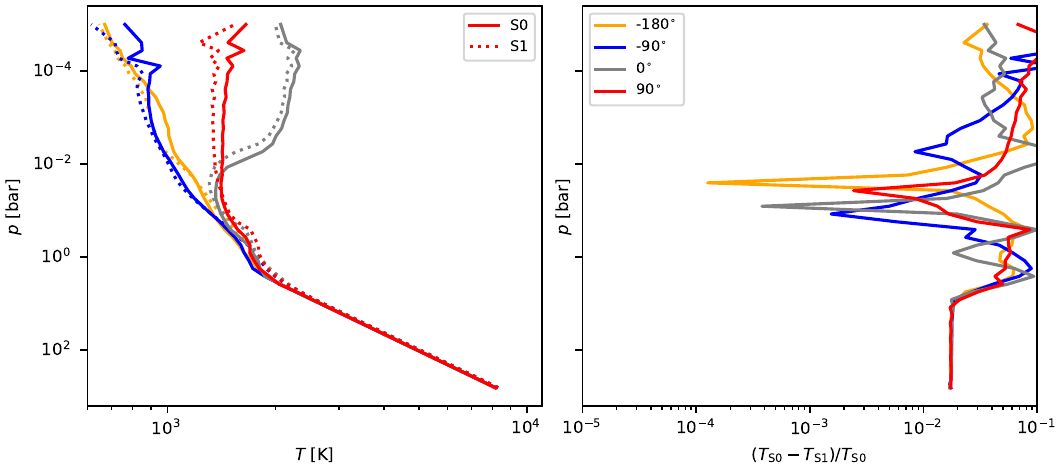}
        \caption{Additional test demonstrating the performance of S0 vs S1 by looking at the final temperature differences (at 12000 days) for several longitudes. The S1 simulation is initialized at 10000 days using the state of the nominal S0 simulation.}
        \label{fig:S0vsS1}
\end{figure*}
We perform an additional simulation of HD~209458b using the S1 resolution in order to benchmark the S0 resolution used in the model (see Table~\ref{tab:opac_bins_overview}). The wavelength resolution S1 is introduced in \citet{Kataria2013}. This additional simulation is initialized with the results of the S0 simulation, shown in Sect.~\ref{sec:HD2} at 10000 days. We then run the model for 2000 days and compare the resulting temperature profile in Fig.~\ref{fig:S0vsS1}.

We find that differences between the two wavelength resolutions are on the order of a few percent in the observable part of the atmosphere, while the relative error decreases with pressure. The hotter temperature of the upper layers in the S0 simulation can partly be accounted for by the stronger incoming flux, as discussed above. We note that the smaller error in the deeper layers of the atmosphere demonstrates that the deep layers are not radiatively driven, and can only be driven by other mechanisms such as dynamics at greater depths.

\subsection{Performance of the interpolation scheme}\label{sec:interp_vs_nointerp}
\begin{figure*}
        \centering
        \includegraphics[width=\textwidth]{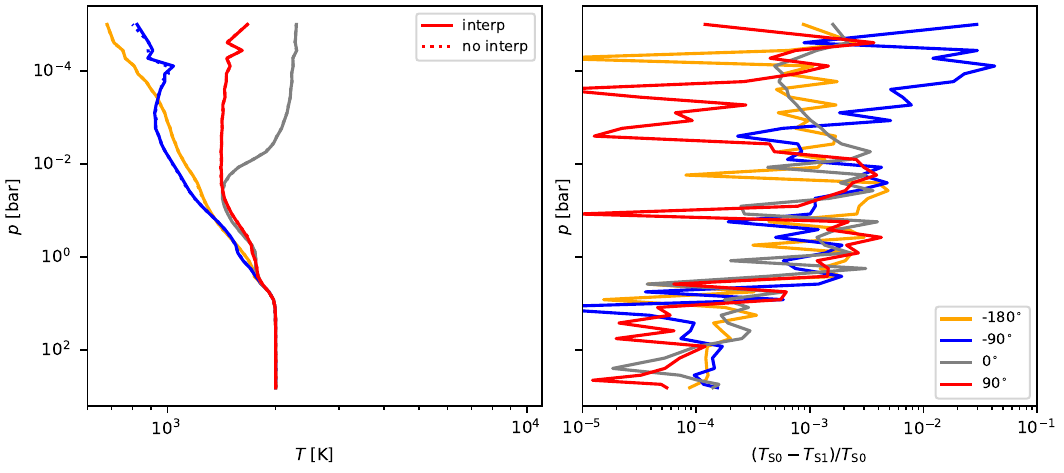}
        \caption{Additional test demonstrating the performance of the flux interpolation scheme (see Fig.~\ref{fig:convergence} and Sect.~\ref{sec:RT_flux}) by looking at an isothermally (T=$\SI{2000}{\K}$) initialized HD~209458b simulation with and without the horizontal interpolation of fluxes.}
        \label{fig:interp_vs_nointerp}
\end{figure*}

We performed an additional short term simulation (run for \num{100} days) without the horizontal interpolation of fluxes, where we solved the radiative transfer equation in every grid cell instead of every second grid cell (see Fig.~\ref{fig:convergence}). These models are identically to the cold HD~209458b model presented in  Sect.~\ref{sec:temp_convergence}, where we use a globally isothermal temperature of \SI{2000}{\K} to initialize the model. We show the resulting time averaged temperature after 100 days of the model with and without interpolation in Fig.~\ref{fig:interp_vs_nointerp} and find that both agree within 2\%. We can therefore conclude that the small wavelength resolution introduces a larger error compared to the error introduced by the horizontal interpolation of the fluxes.

\FloatBarrier
\section{Parameters}
\begin{table*}
        \centering
        \caption{Parameters}
        \begin{tabular}{c c c c}
        \hline\hline
        Parameter & Meaning & HD~209458b & WASP-43b\\
        \hline
        $\Delta t$ & dynamical time-step & \SI{25}{\s} & \SI{25}{\s}\\
        $T_\star$ & stellar temperature & \SI{6092}{\K} & \SI{4520}{\K}\\
        $R_\star$ & stellar radius & \SI{1.203}{R_\odot} & \SI{0.667}{R_\odot}\\
        $a_p$ & semi-major axis & \SI{0.04747}{\AU} & \SI{0.01526}{\AU}\\
        $T_\mathrm{irr}$ & substellar irradiation temperature & \SI{1479}{\K} & \SI{1441}{\K}\\
        $\Theta_\mathrm{ad}$ & initial temperature at \SI{1}{\bar} & \SI{1800}{\K} & \SI{1400}{\K}\\
        $R_p$ & planetary radius & \SI{1.38}{R_\mathrm{Jup}} & \SI{1.036}{R_\mathrm{Jup}} \\
        $c_p$ & specific heat capacity at const. p & \SI{12766}{\J\per\kg\per\K} & \SI{12766}{\J\per\kg\per\K}\\
        $R$ & specific gas constant & \SI{3590}{\J\per\kg\per\K} & \SI{3590}{\J\per\kg\per\K}\\
        $P_\mathrm{rot}$ & rotation period & \SI{3.47}{\day} & \SI{0.8135}{\day}\\
        $g$ & surface gravity & \SI{8.98}{\m\per\s\squared} & \SI{46.9}{\m\per\s\squared}\\
        $k_\mathrm{top}$ & sponge layer Rayleigh friction & \SI{20}{\per\day} & \SI{20}{\per\day}\\
        $\tau_\mathrm{deep}$ & timescale for Rayleigh friction in deep layers & \SI{1}{\day} & \SI{1}{\day}\\

        $p_\mathrm{top}$ & lowest pressure level &\SI{1e-5}{\bar} & \SI{1e-5}{\bar}\\
        $p_\mathrm{bottom}$ & highest pressure level &\SI{700}{\bar} & \SI{700}{\bar}\\
        $N_\mathrm{layers}$ & vertical resolution & 47 & 47 \\
        \hline
        \end{tabular}
        \begin{tablenotes}
                \item \textbf{Notes:} Parameters used in simulations throughout this work.
        \end{tablenotes}

        \label{tab:param}
\end{table*}
\begin{table}
        \centering
        \caption{Radiative time-steps}
                \begin{tabular}{c c c}
                \hline\hline
                time [days] & $\Delta t^\mathrm{rad}_\mathrm{HD2}$ [s] & $\Delta t^\mathrm{rad}_\mathrm{W43b}$ [s]\\
                \hline
                \num{0}-\num{500} & \num{100} & \num{100} \\
                \num{500}-\num{12000} & \num{300} & \num{100} \\
                \end{tabular}
        \begin{tablenotes}
                \item \textbf{Notes:} Radiative time-steps $\Delta t^\mathrm{rad}_\mathrm{HD2}$ and $\Delta t^\mathrm{rad}_\mathrm{W43b}$ for the simulations shown in Sect. \ref{sec:HD2} and \ref{sec:WASP-43b} respectively. The radiative time-step is increased stepwise, since temperature tendencies decrease with time.
        \end{tablenotes}
        \label{tab:radtimesteps}
\end{table}
We show the parameters that we employ in our models of HD~209458b and WASP-43b in Table~\ref{tab:param}. We update the radiative fluxes of WASP-43b more often than we do for HD~209458b, since our models of WASP-43b keep evolving in temperature even after hundreds of days, whereas our model of HD~209458b shows no sign of temperature evolution after a few hundred days. The corresponding numerical flux update rates (radiative time-steps) are shown in Table~\ref{tab:radtimesteps}.

\FloatBarrier
\section{Simplified forcing vs full radiative transfer}\label{sec:newton}

\begin{figure*}
        \centering
        \includegraphics[width=\textwidth]{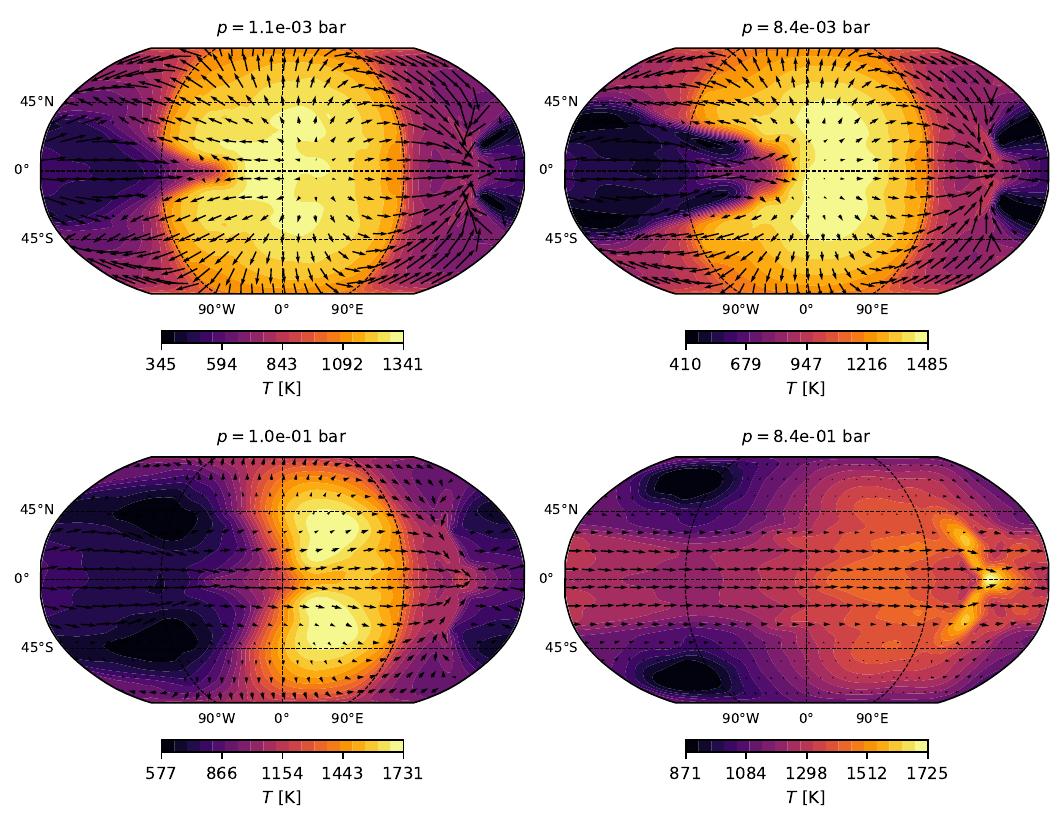}
        \caption{Maps of a model of HD~209458b with Newtonian cooling displaying the color coded temperature at different pressure levels (like Fig.~\ref{fig:HD2_temp}).}
        \label{fig:HD2_temp_robin}
\end{figure*}
\begin{figure}
        \centering
        \includegraphics[width=.45\textwidth]{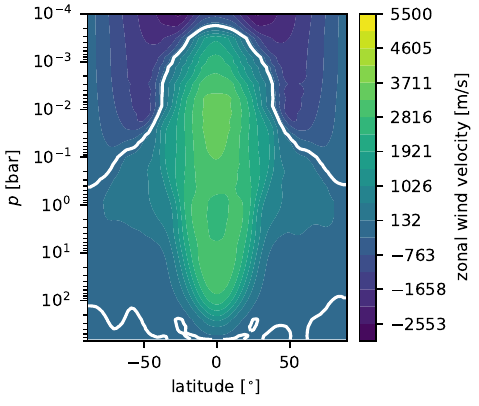}
        \caption{Zonally averaged zonal wind velocities (like Fig. \ref{fig:HD2_zonal_mean}) for the model of HD~209458b with Newtonian cooling.}
        \label{fig:HD2_zonal_mean_robin}
\end{figure}
\begin{figure*}
        \centering
        \includegraphics[width=\textwidth]{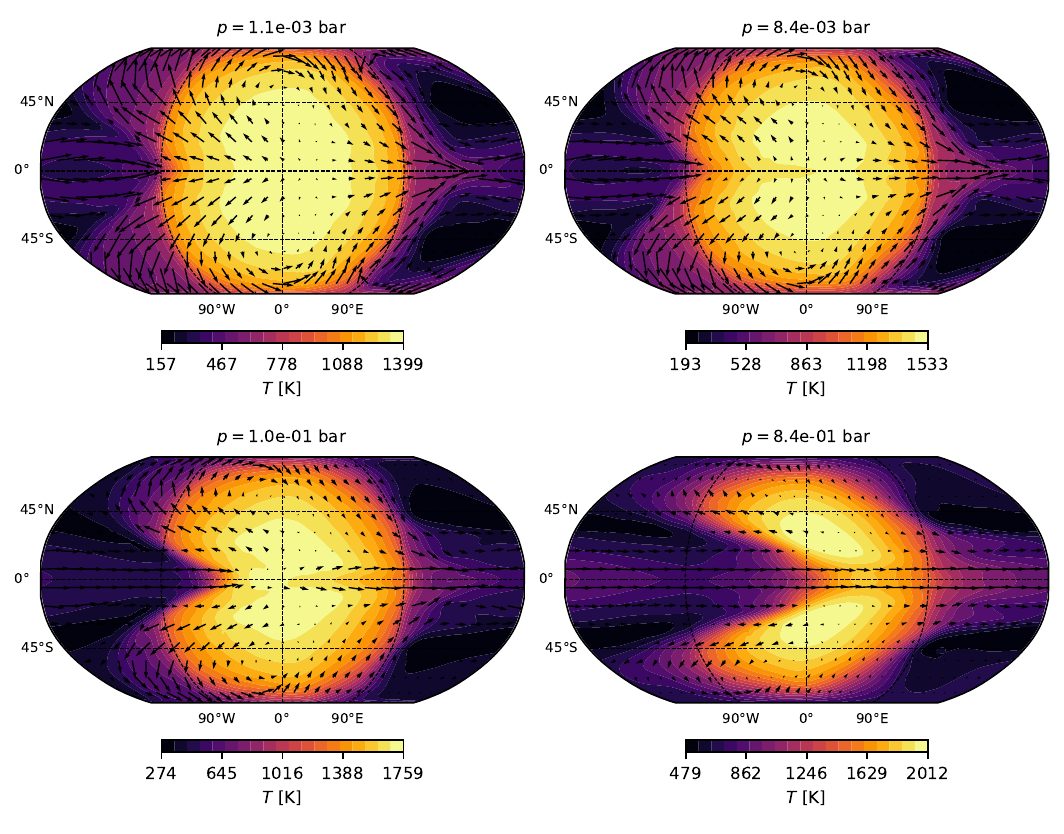}
        \caption{Maps of WASP-43b with Newtonian cooling displaying the color coded temperature at different pressure levels (like Fig.~\ref{fig:WASP-43b_temp}).}
        \label{fig:WASP-43b_temp_robin}
\end{figure*}
\begin{figure}
        \centering
        \includegraphics[width=.45\textwidth]{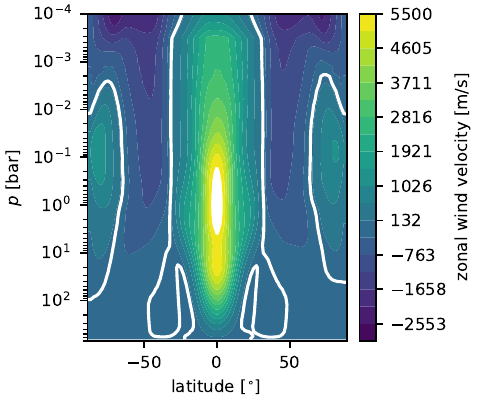}
        \caption{Zonally averaged zonal wind velocities (like Fig. \ref{fig:WASP-43b_zonal_mean}) for the model of WASP-43b with Newtonian cooling.}
        \label{fig:WASP-43b_zonal_mean_robin}
\end{figure}

Our simulations of HD~209458b and WASP-43b with fully coupled radiative feedback are compared to similar simulations from a grid of 3D GCM models taken from \citet{Baeyens2021GCM} with different parameters. The grid models are forced using the Newtonian cooling formalism and use the same model as in \citet{Carone2020GCM}. Specifically, to match the climate properties of HD~209458b a model with $T_\textrm{eff} = \SI{1400}{\K}$, $g = \SI{10}{\m\per\s\squared}$ and host star type G5 is selected (Figs~\ref{fig:HD2_temp_robin} and \ref{fig:HD2_zonal_mean_robin}). For the WASP-43b-like case, we compare our simulation to a model with $T_\textrm{eff} = \SI{1400}{\K}$, $g = \SI{100}{\m\per\s\squared}$ and host star type K5 (Figs~\ref{fig:WASP-43b_temp_robin} and \ref{fig:WASP-43b_zonal_mean_robin}). Additional information about these grid models can be found in \citet{Baeyens2021GCM}. Because the Newtonian-cooled grid models and the fully coupled models in this work make use of the same dynamical core (\texttt{MITgcm}) and similar radiative transfer module (\texttt{petitRADTRANS}), their comparison highlights the important role of radiative feedback in hot Jupiter climates. We note that the grid models do not use TiO and VO opacities, while the fully coupled models do include TiO and VO, leading to a temperature inversion in the fully coupled models, which is not to be expected in the Newtonian-cooled models.

In general, the Newtonian-cooled grid models show qualitative agreement with the simulations presented in this work, but quantitative differences are present. For the case of HD~209458b, the grid model (Fig.~\ref{fig:HD2_temp_robin}) appears to be cooler than the fully coupled model (Fig.~\ref{fig:HD2_temp}), in particular regarding the nightside of the planet, owing to the colder effective temperature of the grid model. The equatorial jet is not as fast in the grid model as in the fully coupled simulation. In the former, it reaches wind speeds up to \SI{5.0}{\km\per\s}, whereas in the latter, wind speeds up to \SI{7.2}{\km\per\s} are attained (see Figs~\ref{fig:HD2_zonal_mean_robin} and \ref{fig:HD2_zonal_mean}). Finally, the jet persists up to the lowest pressures in the fully coupled simulation, whereas in the Newtonian-cooled model, the low-pressure regime is characterized by a thermally direct day-to-night circulation. The transition between a shallow and a deep jet seems to be for rotation periods between two and three days \citep[e.g, compare first two columns of Fig.~3 from][]{Baeyens2021GCM}, and because the grid model is not a perfect match to HD 209458b, it falls on the different side of this transition.

The disappearance of cold nightsides with a reduced day-night temperature contrast and a thermally direct wind flow at high altitudes, when full radiative feedback is taken into account, can also be seen in \citet{Amundsen2016UKMetGCM}. A more efficient day-night heat redistribution, when radiative feedback is taken into account, was also noted by \citet{Showman2009GCM}. In the models of \citet{Amundsen2016UKMetGCM}, however, the wind speed of the zonal jet stream is slightly slower in the fully coupled GCM than in the Newtonian-cooled GCM. In contrast, we find that the full radiative feedback yields a zonal jet stream that is slightly faster than the Newtonian-cooled case. Hence, further detailed comparisons between the two forcing mechanisms are necessary to establish their impact on the jet stream speed.

Finally, for the case of WASP-43b, the same quantitative changes discussed above can be observed when a Newtonian-cooled and fully coupled model are compared. Additionally, a climate transition is observed, in which the Newtonian-cooled model yields equatorial retrograde flow (Fig.~\ref{fig:WASP-43b_temp_robin}), but the fully coupled model does not (Fig.~\ref{fig:WASP-43b_temp}). A possible explanation for the disappearance of retrograde flow in the latter case could be that retrograde and superrotating equatorial flow are competing tendencies \citep{Carone2020GCM}. The increase in jet strength in the fully coupled model, as established in the HD~209458b comparison in the previous paragraph, would then become the dominant tendency in WASP-43b. Given that the fully coupled model should be the more realistic of the two forcing mechanisms, future investigations will be needed to investigate whether retrograde equatorial wind flow can also be elicited in models with radiative feedback.

\end{appendix}
\end{document}